\documentclass[12pt,preprint]{aastex}

\newcommand{\be}{\begin{equation}}
\newcommand{\ee}{\end{equation}}\newcommand{\beq}{\begin{eqnarray}}
\newcommand{\eeq}{\end{eqnarray}}
\newcommand\kms   {km~s$^{-1}$}
\newcommand\accu   {m~s$^{-2}$}

\newcommand{\halpha}{\mbox{H\hspace{0.2ex}$\alpha$}}

\slugcomment{to be submitted to ApJ}

\shortauthors{De Pontieu, et al. }
\shorttitle{Observations and Modeling of Fibrils}

\begin{document}
\title{High Resolution Observations and Modeling of Dynamic Fibrils}

\author{B. De Pontieu}
\affil{Lockheed Martin Solar and Astrophysics Lab, 3251 Hanover St.,
  Org. ADBS, Bldg. 252, Palo Alto, CA 94304, USA}
\email{bdp@lmsal.com}
\author{V.H. Hansteen\altaffilmark{1}}
\affil{Institute of Theoretical Astrophysics, University of Oslo, PO
  Box 1029 Blindern, 0315 Oslo, Norway}
\email{viggo.hansteen@astro.uio.no}
\author{L. Rouppe van der Voort\altaffilmark{1}}
\affil{Institute of Theoretical Astrophysics, University of Oslo, PO
  Box 1029 Blindern, 0315 Oslo, Norway}
\email{rouppe@astro.uio.no}
\author{M. van Noort\altaffilmark{2}}
\affil{Institute of Theoretical Astrophysics, University of Oslo, PO
  Box 1029 Blindern, 0315 Oslo, Norway}
\altaffiltext{2}{Now at: Institute for Solar Physics of the Royal 
Swedish Academy of Sciences, AlbaNova University Center, 106 91 Stockholm, Sweden}
\email{noort@astro.su.se}
\author{M. Carlsson\altaffilmark{1}}
\affil{Institute of Theoretical Astrophysics, University of Oslo, PO
  Box 1029 Blindern, 0315 Oslo, Norway}
\altaffiltext{1}{Also at: Center of Mathematics for Applications, 
University of Oslo, P.O.~Box~1053, Blindern, N--0316 Oslo, Norway}
\email{mats.carlsson@astro.uio.no}
\begin{abstract}
  We present unprecedented high resolution H$\alpha$ observations,
  obtained with the Swedish 1-m Solar Telescope, that, for the first
  time, spatially and temporally resolve dynamic fibrils in active
  regions on the Sun.  These jet-like features are similar to mottles
  or spicules in quiet Sun. We find that most of these fibrils follow
  almost perfect parabolic paths in their ascent and descent.  We
  measure the properties of the parabolic paths taken by 257 fibrils,
  and present an overview of the deceleration, maximum velocity,
  maximum length and duration, as well as their widths and the
  thickness of a bright ring that often occurs above dynamic fibrils.
  We find that the observed deceleration of the projected path is
  typically only a fraction of solar gravity, and incompatible with a
  ballistic path at solar gravity. We report on significant
  differences of fibril properties between those occurring above a
  dense plage region, and those above a less dense plage region where
  the magnetic field seems more inclined from the vertical.  We
  compare these findings to advanced numerical 2D radiative MHD
  simulations, and find that fibrils are most likely formed by
  chromospheric shock waves that occur when convective flows and
  global oscillations leak into the chromosphere along the field lines
  of magnetic flux concentrations. Detailed comparison of observed and
  simulated fibril properties shows striking similarities of the
  values for deceleration, maximum velocity, maximum length and
  duration.  We compare our results with observations of mottles and
  find that a similar mechanism is most likely at work in the quiet
  Sun.
\end{abstract}

\keywords{magnetic fields --- Sun: photosphere --- Sun: chromosphere}

\section{Introduction}
The chromosphere is in a highly dynamic state, varying on timescales
of minutes or less. Much of the dynamics in the magnetized regions
associated with the magnetic network and plage are dominated by
short-lived, jet-like features. A whole range of names has been
applied to describe these chromospheric features. At the quiet Sun
limb, they are traditionally called spicules, where they are observed
in \halpha\ as thin, elongated features that develop speeds of 10-30
\kms\ and reach heights of on average 5-9 Mm during their lifetimes of
3 to 15 minutes \citep{Beckers1968}. Their widths (0.2-1 Mm) and
dynamics have, until recently, been very close to the resolution
limits of observations. As a result of this limitation, the
superposition inherent in limb observations, and the difficulty
associated with interpreting H$\alpha$ data, the properties of
spicules have not been very well constrained, which has led to a
multitude of theoretical models \citep[for a review,
see,][]{Sterling2000}.  The same issues have plagued a solid
identification of a counterpart to spicules on the disk. Observations
on the disk in H$\alpha$ reveal dark mottles in quiet Sun regions.
These dark mottles seem similar to spicules, and usually appear in
close association with small flux concentrations (and bright points)
in the magnetic network that outlines the supergranular cells. There
has been a long-standing discussion on whether spicules and mottles
are the same phenomena.  This discussion has focused on the different
velocity distributions of mottles and spicules to conclude that mottles
are not the disk counterpart of spicules
\citep{Grossmann-Doerth+Schmidt1992}.  However, several groups have
used more recent data to argue for a close relationship between
mottles and spicules \citep[see,
e.g.,][]{Tsiropoula+etal1994,Suematsu+etal1995,Christopoulou+etal2001}.

Dynamic, jet-like features can also be seen in and around active
region plage regions.  These features are shorter (1-4 Mm) and
shorter-lived (3-6 minutes) and appear to form a subset of what have
traditionally been called active region fibrils. There are also
fibrils that do not show jet-like behavior (Fig.~\ref{fig_context}).
These are apparently low-lying, heavily inclined and more static
loop-like structures that connect plage regions with opposite polarity
magnetic flux. In the remainder of this paper, we will not discuss
these low-lying fibrils, but focus on the motions of the shorter
dynamic fibrils (hereafter abbreviated as DFs).

Only recently have DFs become the focus of detailed study, with
several papers studying their dynamics, influence on the transition
region and their oscillatory properties
\citep{DePontieu+etal1999,DePontieu+etal2003,DePontieu+etal2003b,DePontieu+etal2004,DePontieu+etal2005,Tziotziou+etal2004}.
In the past, several authors have reported periodicities and
oscillations in spicule observations, usually with dominant periods of
order 5 minutes. Such periodicities in radial velocity, half-width and
line intensity have been suggested based on older observations at the
limb, but their interpretation has been difficult due to the
complicated line formation of H$\alpha$, line-of-sight superposition
and limited spatial resolution
\citep{Beckers1968,Platov+Shilova1971,Kulidzanishvili+Nikolsky1978}.
More recently, \citet{DePontieu+etal2003b} have discovered significant
oscillatory power in DFs in active regions using high-resolution data
from the Swedish 1~m Solar Telescope \citep{Scharmer2003SST}. Not all
DFs show oscillations or recurrence, but a significant number do.
Most of the power in especially more inclined DFs resides in periods
between 4 and 6 minutes
\citep{DePontieu+etal2004,DePontieu+Erdelyi2006}.  Similar periods
have been found by \citet{Tziotziou+etal2004} in mottles in enhanced
network or weak plage. \citet{DePontieu+etal2004} and
\citet{DePontieu+Erdelyi2006} developed a new numerical model to
explain the cause of these (quasi)periodic mottles or DFs. They
compared their model to data from the Transition Region and Coronal
Explorer \citep[TRACE][]{Handy+etal1999} and suggested that inclined
magnetic fields allow leakage of normally evanescent photospheric
p-modes (which have dominant periods of 5 minutes) into the atmosphere
where they form shocks that drive DFs. A recent paper by
\citet{Hansteen+etal2006} uses extremely high resolution observations
obtained at the SST and significantly more advanced numerical
modelling to show that dynamic fibrils are indeed caused by
magneto-acoustic shock waves.

In this paper, we directly expand on the work by
\citet{Hansteen+etal2006}. Here we perform a more comprehensive
analysis of the properties and motion of DFs using the same extremely
high spatial and temporal resolution SST dataset of
\citet{Hansteen+etal2006}. In addition, we use advanced 2/3D numerical
radiative MHD simulations that include the convection zone,
photosphere, chromosphere and lower corona to interpret all of the
measurements and correlations found in the SST data analysis, many of
which have not been described in \citet{Hansteen+etal2006}. In
\S~\ref{sec:instrumentation} we discuss the details of the
observations and instrumentation, including a description of a novel,
advanced optical setup that allowed a diffraction limited 88 minute
long timeseries in \halpha\ linecenter.  We describe the typical
temporal evolution and measure properties of DFs in
\S~\ref{section:DFdescription}. This section also deals with
intriguing regional differences and correlations in DF properties that
seem to be related to the oscillatory behavior of DFs. To interpret
these findings, we use a numerical model that is described in
\S~\ref{section:discussion}. This section also summarizes our
observational findings and contains a more detailed description of the
new model \citep[introduced by][]{Hansteen+etal2006} that explains our
observations. Our results strongly suggest that DFs naturally form as
the result of chromospheric shock waves driven by convective flows and
(global) oscillations in the photosphere as suggested by
\citet{DePontieu+etal2004}.

In \S~\ref{section:qs} we discuss in more detail how this model and
our observations of active region DFs relate to equivalent quiet Sun
phenomena.  The dynamics of quiet Sun mottles have been studied
extensively by several groups recently
\citep{Tsiropoula+etal1994,Suematsu+etal1995,Christopoulou+etal2001,Tziotziou+etal2003,Al+etal2004,Tsiropoula+Tziotziou2004,Tziotziou+etal2004}.
While the lack of high spatial and temporal resolution data has made
it difficult to properly interpret the findings, this recent work
seems to have resolved a long-standing issue \citep{Nishikawa1988}
regarding the paths mottles follow during their ascent and descent.
\citet{Suematsu+etal1995} and \citet{Christopoulou+etal2001} use
\halpha\ data to establish that mottles follow parabolic paths. In
fact, observations by \citet{Christopoulou+etal2001} indicate that
spicules at the limb similarly trace parabolic paths. Mottles
initially appear in the blue wing of \halpha\, extending in the
process to lengths of some 2 to 8 Mm \citep{Suematsu+etal1995}.  After
reaching maximum extent, they fade from the blue wing and appear in
the red wing of \halpha\ to recede generally along the same path.
Throughout most of their lifetime they appear as dark, elongated
features in \halpha\ linecenter. The measurements of the parabolic
paths traced by mottles have been difficult to interpret. The
projected decelerations of the top of the mottle are generally
observed to be far less than solar gravity.  Attempts have been made
to interpret such low decelerations as projection effects superposed
on ballistic flights.  However, the projection angles necessary are
typically not compatible with measured Doppler shifts
\citep{Suematsu+etal1995}, ruling out a purely ballistic
interpretation. In addition, a purely ballistic flight would imply
large initial velocities that have not been observed
\citep{Christopoulou+etal2001,Suematsu+etal1995}. These observational
findings have been difficult to reconcile with the plethora of models
for spicules or mottles \citep{Sterling2000}. In \S~\ref{section:qs}
we describe the similarities between mottles and DFs in more detail,
and discuss the applicability of our model to quiet Sun mottles.  Our
results are summarized in \S~\ref{section:summary}.

\section{Observations and Instrumentation}
\label{sec:instrumentation}

The observations were obtained with the Swedish 1-m Solar Telesope
on La Palma. The Solar Optical
Universal Polarimeter \citep[SOUP, ][]{title81SOUP} was used to obtain
narrow-band images in the linecenter of \halpha\ (filter FWHM
12.8~pm). To compensate for seeing deformations in real-time, the
observations were aided by the use of the SST adaptive optics system
\citep[AO,][]{Scharmer2003AO} consisting of a tip-tilt mirror, a
bimorph mirror with 37 actuators, a correlation tracker, and a
Shack--Hartmann wavefront sensor. As a further measure to reduce
seeing aberrations, the data were post-processed with the Multi-Object
Multi-Frame Blind Deconvolution image restoration method
\citep[MOMFBD, ][]{vanNoort05MOMFBD}. This method is complementary to
the use of AO since it reduces residual aberrations not corrected by
the AO and it corrects the areas in the FOV that are not used for the
AO wavefront sensing.

\begin{figure}
\epsscale{1}
\plotone{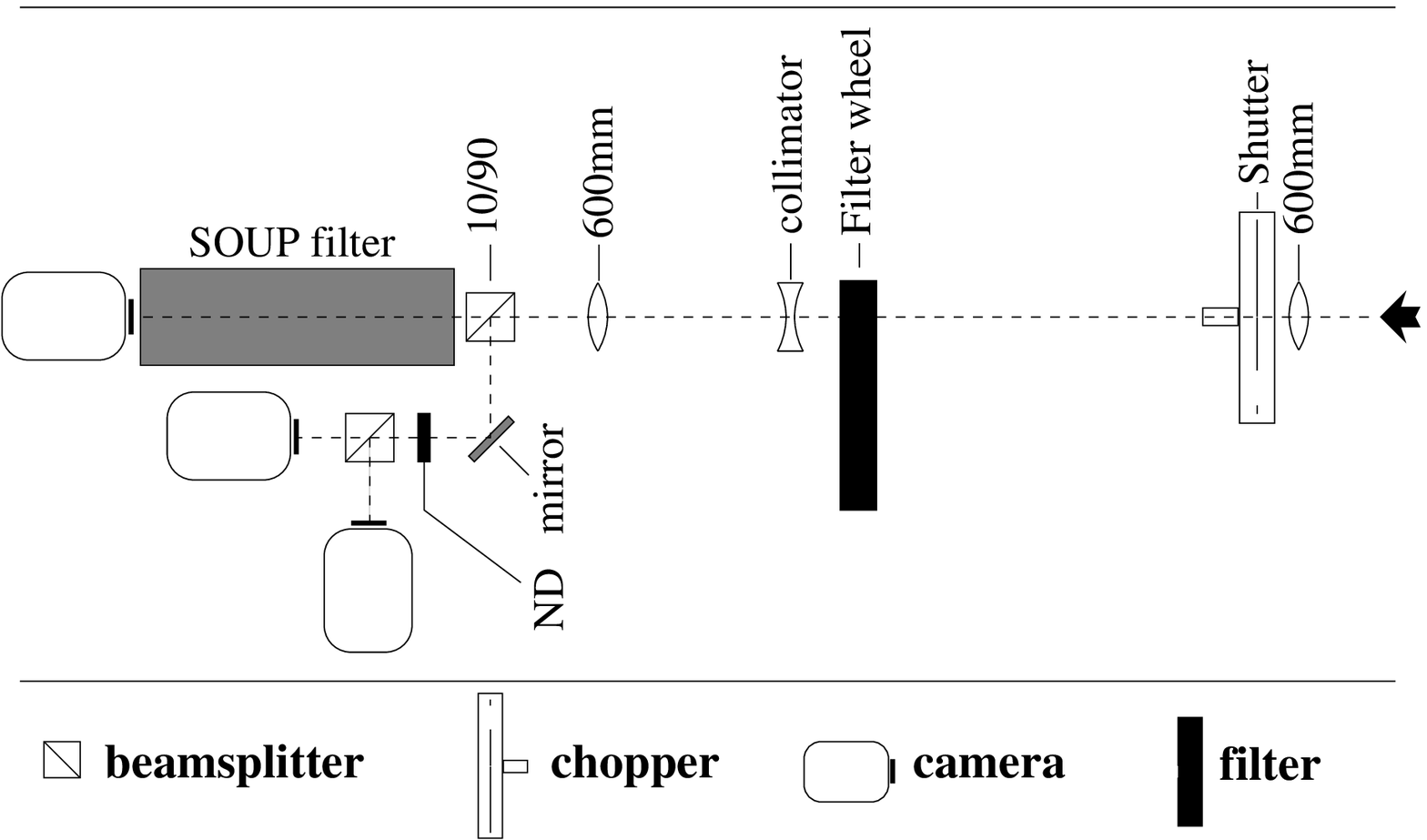}
\caption{Optical setup at the Swedish 1 m Solar Telescope (SST) in La Palma used
  to obtain \halpha\ data on 4-Oct-2005. See \S
  \ref{sec:instrumentation} for details.
  \label{fig_optical_setup}}
\end{figure}

The science cameras consisted of 3 new, high speed cameras of type
Sarnoff CAM1M100, with a CCD readout time of only 10~ms and an
effective exposure time of 15~ms, resulting in a frame rate of
approximately 37 frames-per-second (fps).  The setup used was
specially built for MOMFBD processing of the data and consisted of one
camera behind the SOUP filter and 2 "wideband" cameras in front of
the SOUP filter but after the SOUP pre-filter (a 0.8~nm FWHM
interference filter centered at 656.3~nm). The light for the
wideband cameras was split off the main beam by a 90/10
beamsplitter in front of the SOUP filter and split again evenly for
the 2 wideband cameras, which were set up as a phase diversity pair
with one camera at 12~mm defocus on a 45~m focal length. A schematic
drawing of the red beam optical set-up is shown in
Fig.~\ref{fig_optical_setup}.

To ensure simultaneous exposure, a specially designed chopper with a
useful aperture of 22$\times$22~mm and a variable duty cycle of
$\frac{1}{2}$ - $\frac{3}{4}$ was used as a shutter. To guarantee
correct labeling of all camera frames, a unique exposure number was
generated by the shutter electronics and stored on the camera PCs.
This scheme guarantees that simultaneously exposed frames can be
uniquely identified, a feature that is required for MOMFBD processing
but difficult to achieve using the internal clock of the observing PCs
at the frame rates used.

The large data rate produced by the cameras (72~MB$\,$s$^{-1}$ per
camera) was reduced by a factor of 3 using real-time data compression,
sent from the camera computers to disk-writing computers through
ordinary gigabit network connections, and then passed through
fiberchannel links to the RAID controllers of an ATABeast system. More
details of the data acquisition, as well as first results from this
unique dataset are described by \citet{vanNoort+Rouppe2006}.

The observations were obtained on 04-Oct-2005 and consist of a time
series of about 78 minute duration, centered on one of the larger
spots in the young active region AR10813 at S7,E37. The observing
angle $\theta$ between the line-of-sight and the local vertical is
$39\degr$ ($\mu=\cos(\theta)=0.78$). The field-of-view (FOV) was
about 64\arcsec$\times$64\arcsec\ at a pixel scale of 0\farcs063. This
corresponds to a FOV of 46,400 x 46,400 km, with a pixel scale of 45.7
km. The diffraction limit of the SST at 6563 \AA\ is 0.165\arcsec,
i.e., 120 km.

\subsection{Data reduction}
\label{sec:reduction}

The images from all cameras were jointly processed with the MOMFBD
image restoration method, yielding 2 restored images for each
restoration (1 wide band and 1 \halpha).  Since the common
field-of-view of the cameras within an MOMFBD set is separately
aligned to sub-pixel accuracy, the restored images have near-perfect
alignment.

Sets from 12 exposures were used per restoration (i.e., 36 images in
total) resulting in a time series with 3~fps. For this study, we
selected every third image, giving a time series with a fixed cadence
of 1~s between images.

The time series were corrected for diurnal field rotation, aligned and
de-stretched. The (local) offsets were determined on the wide band
images and subsequently applied to the \halpha\ images.

\subsection{Description of field of view}

\begin{figure}
\epsscale{1}
\plotone{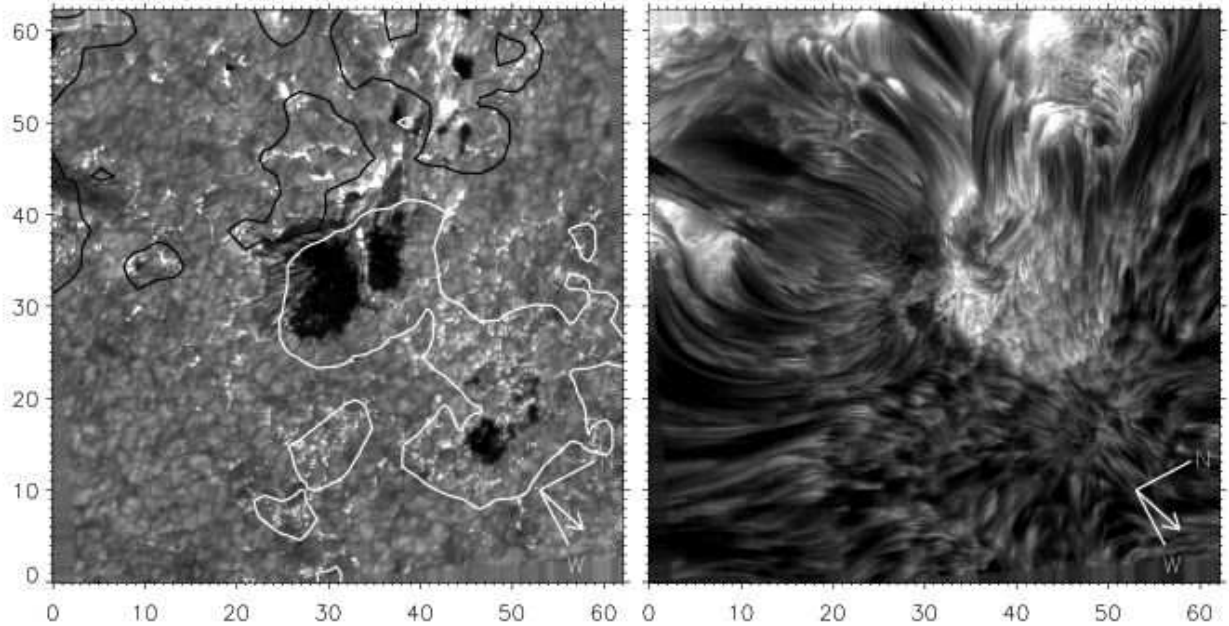}
\caption{Images taken in \halpha\ wideband (left panel) and \halpha\
  linecenter (right panel) of part of NOAA AR 10813 taken at SST on
  4-Oct-2005. Tickmarks are in arcseconds. The dynamic fibrils (DFs)
  are predominantly observed in the mostly unipolar plage region
  between the two sunspots, for x=35-60\arcsec and y=20-40\arcsec. The
  white and black contours in the left panel outline the positive and
  negative magnetic flux from a full-disk MDI magnetogram. The active
  region is located at heliocentric coordinates S7, E37. The direction
  of disk center is indicated by a white straight arrow. An mpeg-movie
  showing the temporal evolution of dynamic fibrils (DFs) in \halpha\ 
  linecenter is available online.
  \label{fig_context}}
\end{figure}

The field of view of the time series contains two small sunspots of
NOAA active region (AR) 10813. The larger of the spots is split into
two parts by a lightbridge as seen in the left panel of
Fig.~\ref{fig_context}. The spots are surrounded by areas of plage,
which includes some pores to the north of the larger spot. The
chromosphere, as imaged in H$\alpha$ linecenter (right panel of
Fig.~\ref{fig_context}), shows many almost horizontal fibrils that
connect the spots to surrounding opposite polarity pores and plage.
Many of these horizontal fibrils are associated with the superpenumbra
of the larger spot. These fibrils are longer (up to 20-40\arcsec) and
more static than the dynamic fibrils that are the focus of this paper.

Most of the DFs are associated with the plage in the
region outlined by x=35-60\arcsec and y=20-45\arcsec. The chromosphere
in this region is dominated by relatively short jet-like features that
appear to rise and fall within a matter of minutes. During their
ascent and descent, individual DFs seem to follow the same, roughly
straight, trajectory. At any given time, scores of DFs are visible in
this region of interest. DFs often seem to occur in semi-coherent
patches which contain several DFs at any given time. The location and
orientation of these patches seems to be governed by the topology of
the underlying magnetic plage region. Within such a patch, it is
sometimes difficult to uniquely identify individual DFs because of the
semi-coherent behavior over a few arcseconds: some of these coherent
fronts show significant substructure.

The region which is dominated by DFs appears somewhat brighter in
H$\alpha$ linecenter than surrounding regions. This may be because hot
coronal loops are connected to this region. Such a correlation has
previously been described by \citet{DePontieu+etal2003} who find a correlation
between the brightness in H$\alpha$ linecenter and that of upper
transition region moss emission in Fe IX/X 171 \AA\ (using TRACE
data). Bright moss emission occurs at the footpoint regions of hot
($>3$ MK) coronal loops.

\begin{figure}
\epsscale{1}
\plottwo{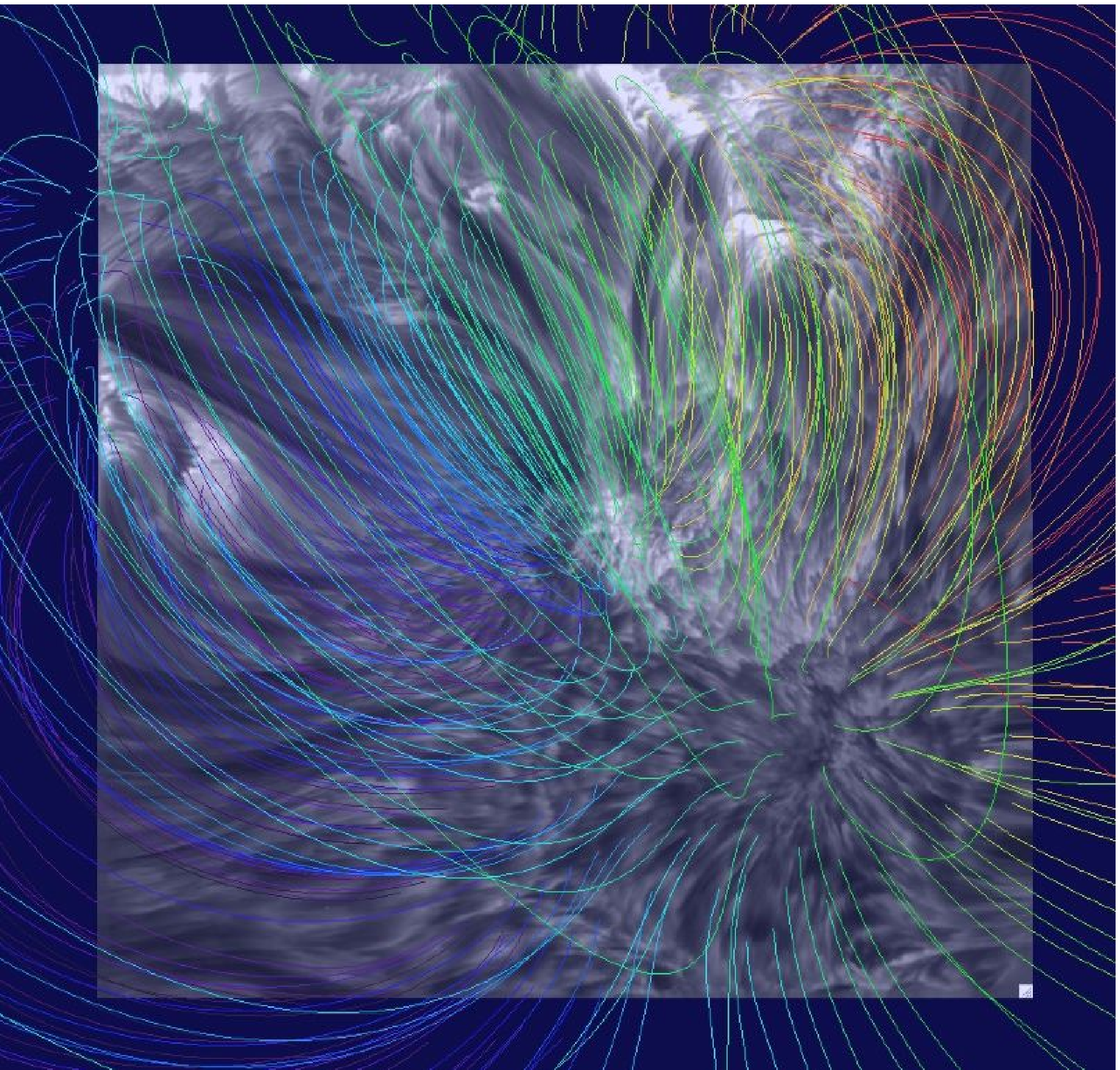}{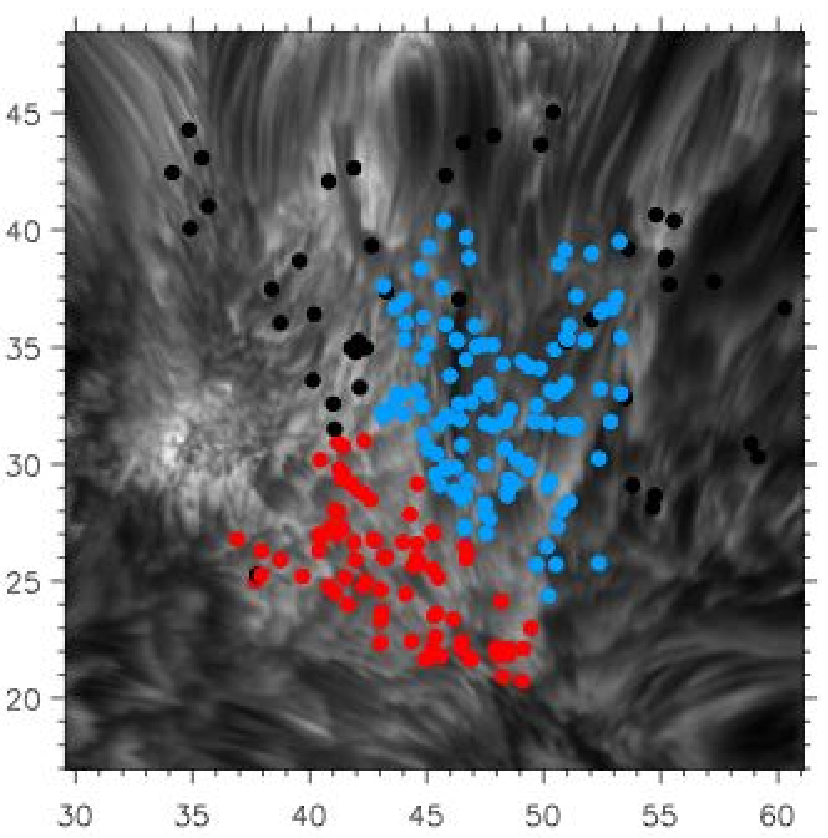}
\caption{Left panel shows a potential field extrapolation superposed
  on a \halpha\ linecenter image. The potential field extrapolation is
  based on a full-disk MDI magnetogram in the middle of the time
  series. Colors of fieldlines are only chosen for clarity. Right
  panel shows a \halpha\ linecenter image with positions of DFs
  overplotted. Blue dots indicate DFs located in region 1, whereas red
  dots indicate DFs located in region 2. All other DFs that were
  measured are shown as black dots. Region 2 is the site of a dense
  plage area where magnetic fields are more concentrated and more
  vertical. Region 1 is adjacent to this dense plage region. Magnetic
  fields here are more inclined as the field connects to the opposite
  polarity plage that surrounds the pores at the top of Fig.~\ref{fig_context}.
  \label{fig_potext}}
\end{figure}

A full disk magnetogram taken by SOHO/MDI, centered in the middle of
the time series, at 09:39 UT on 4-Oct-2005, shows that the two spots
and plage in the region of interest share the same polarity (see
contours of Fig.~\ref{fig_context}). Most of the associated magnetic
flux seems to connect to opposite polarity to the north and west of
the spots, for example the pores to the north and the plage outside the
field of view to the north. This can be deduced from the potential
field extrapolation shown in the left panel of Fig.~\ref{fig_potext}.
The field extrapolation seems to correlate reasonably well with
features outlined by H$\alpha$ in some regions, including our region
of interest. However, the extrapolation deviates from the H$\alpha$
observations in several regions of the active region to the north and
east of the spots, as well as close to the pores to the north of the
large sunspot.

Some DF properties vary significantly with position in the field of
view. Several regions were defined: region 1 is at the edge of a plage
region where the magnetic field seems more inclined (see discussion
below). Region 2 covers a dense plage region (adjacent to region 1)
where the fields seem more vertical. The average magnetic field
strength, as derived from a full-disk MDI magnetogram (with 2\arcsec\
pixels), is of order 50-100 Mx cm$^{-2}$ for region 1, and 150-200 Mx
cm$^{-2}$ for region 2. The other regions contain fewer DFs and
their magnetic topology is not as well defined as in regions 1 and 2.
The right panel of Fig.~\ref{fig_potext} illustrates the location of
the two main regions (blue is region 1, red is region 2).

\section{Description of Dynamic Fibrils}
\label{section:DFdescription}
\subsection{The life of a fibril}

Through visual inspection of a time lapse movie of the \halpha\
linecenter images, we found many fibrils in the plage region in
proximity to the two spots near the center of the image.
We analyzed DFs that occurred throughout the 78 minute long time
series.  To analyze the properties of the DFs, we manually choose the
general direction of a DF (or set of DFs). A guideline is drawn in
this direction, and a subset of the data, parallel to this guideline,
is extracted for further analysis. This procedure results in detailed
movies of individual DFs, as well as 'xt'-plots that show the
evolution of the extent of the DF along the guideline as a function of
time. A total of 257 DFs were analyzed in this fashion.

\begin{figure}
\epsscale{1}
\plotone{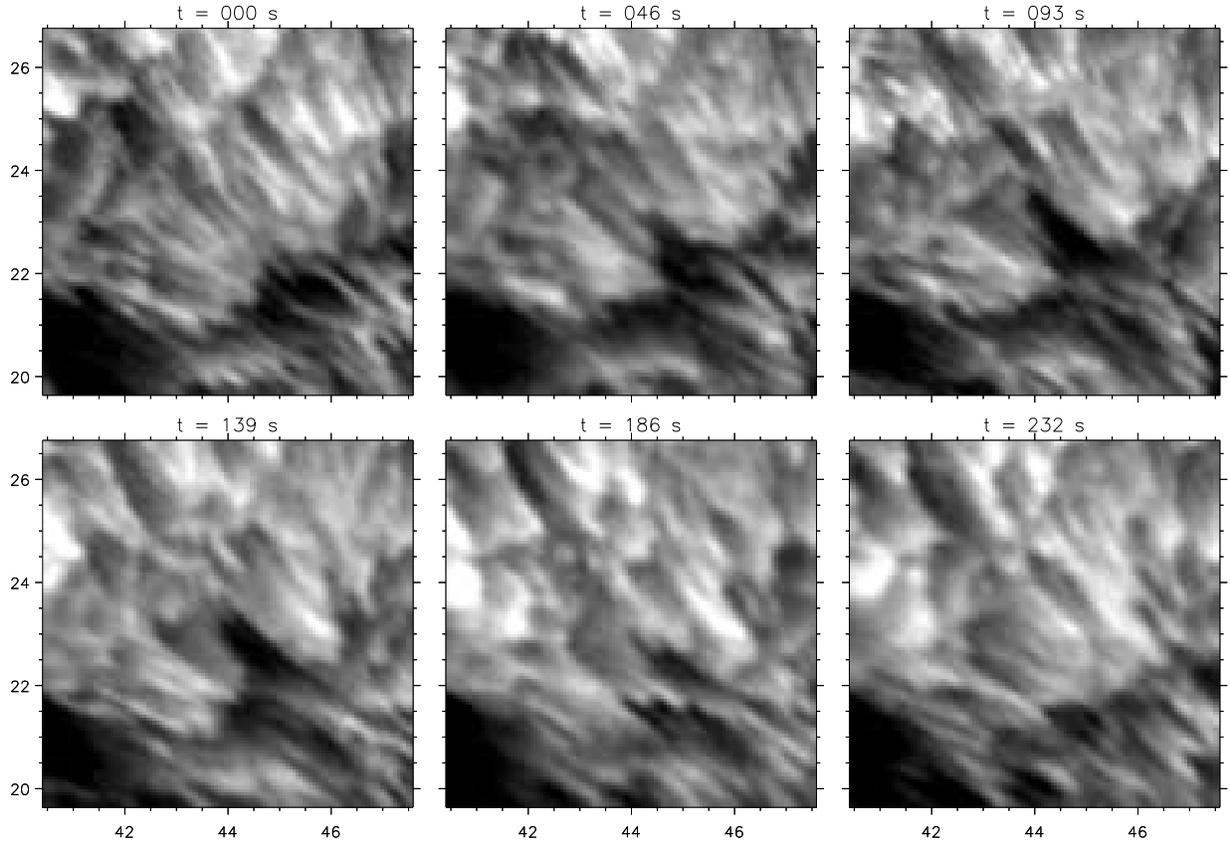}
\caption{Snapshots in \halpha\ linecenter that show the temporal
  evolution of a dynamic fibril as it rises and retreats. This DF is
  thin, short and highly dynamic. It shows some evidence of
  substructure during its evolution. It is located in region 2. An
  mpeg-movie showing the temporal evolution of this DF is available
  online.
  \label{fig_spiclife_1}}
\end{figure}

\begin{figure}
\epsscale{1}
\plotone{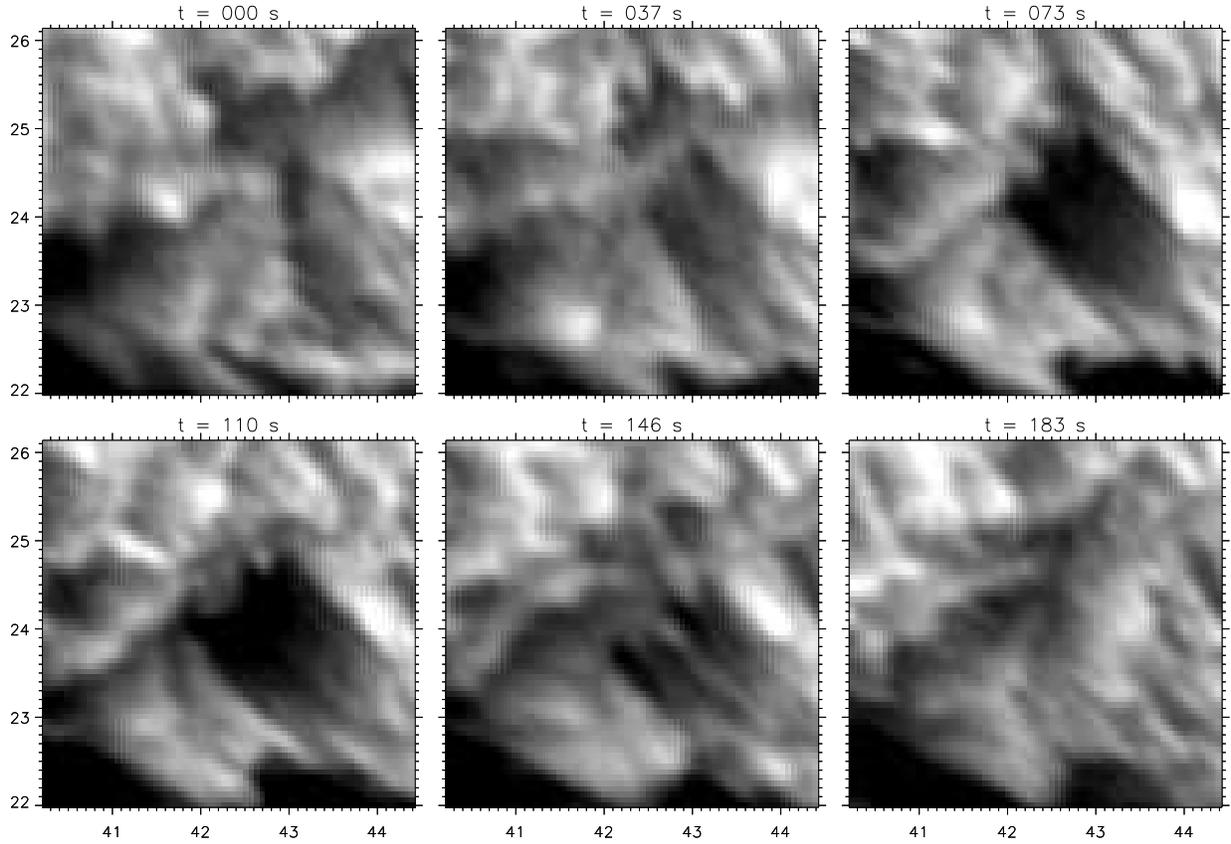}
\caption{Snapshots in \halpha\ linecenter that show the temporal
  evolution of a dynamic fibril as it rises and retreats. This DF is
  relatively wide, short and lives for only three minutes. It is
  located in region 2. It shows clear evidence of substructure during
  its evolution, with different 'fingers' developing in a different
  fashion as a function of time. An mpeg-movie showing the temporal
  evolution of this DF is available online.
  \label{fig_spiclife_2}}
\end{figure}

\begin{figure}
\epsscale{1}
\plotone{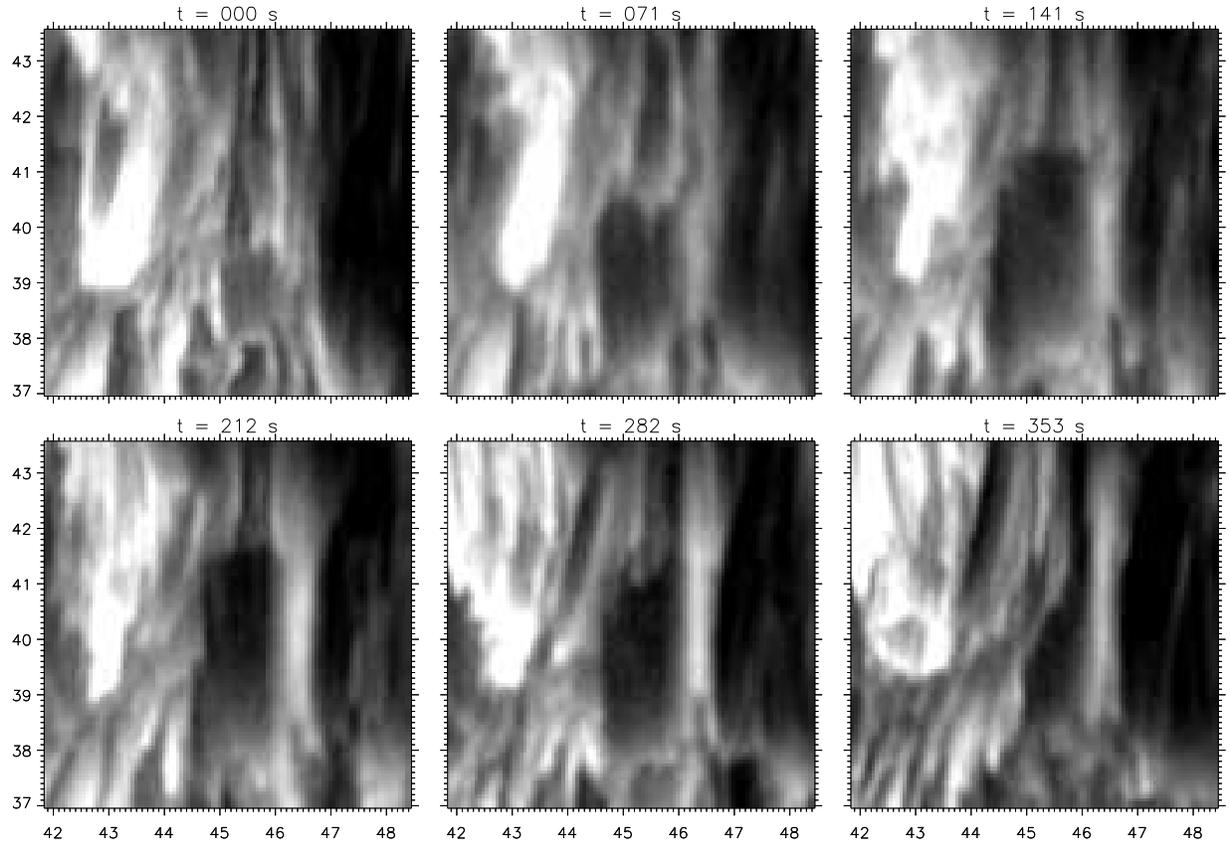}
\caption{Snapshots in \halpha\ linecenter that show the temporal
  evolution of a dynamic fibril as it rises and retreats. This DF is
  very wide, long and relatively slow in its evolution. There are
  actually two DFs occurring along the same line-of-sight, with
  different temporal evolution, as evidenced in
  Fig.~\ref{fig_spicxt}. An mpeg-movie showing the temporal
  evolution of this DF is available online.
  \label{fig_spiclife_3}}
\end{figure}

In Figs.~\ref{fig_spiclife_1}, \ref{fig_spiclife_2}, and
\ref{fig_spiclife_3} we present the evolution of three typical DFs.
Figure~\ref{fig_spiclife_1} shows a DF in region 2. The six panels show that the
DF rises rapidly to a maximum length in roughly 130~s and recedes
along the same path some 260~s after first becoming visible. It is
darker than its surroundings, especially at maximum extension, and
seems to follow a straight path in the direction that is shared by all
surrounding features. The DF has some internal structure, most clearly
visible at $t=139$~s and $t=186$~s. The maximum extent is relatively
modest at some 2\arcsec.  Despite the substructure, the DF is thin
with a width of some $0.5$\arcsec. It is also well isolated even
though a number of other DFs are visible in the immediate vicinity.

\begin{figure}
\epsscale{0.48}
\plotone{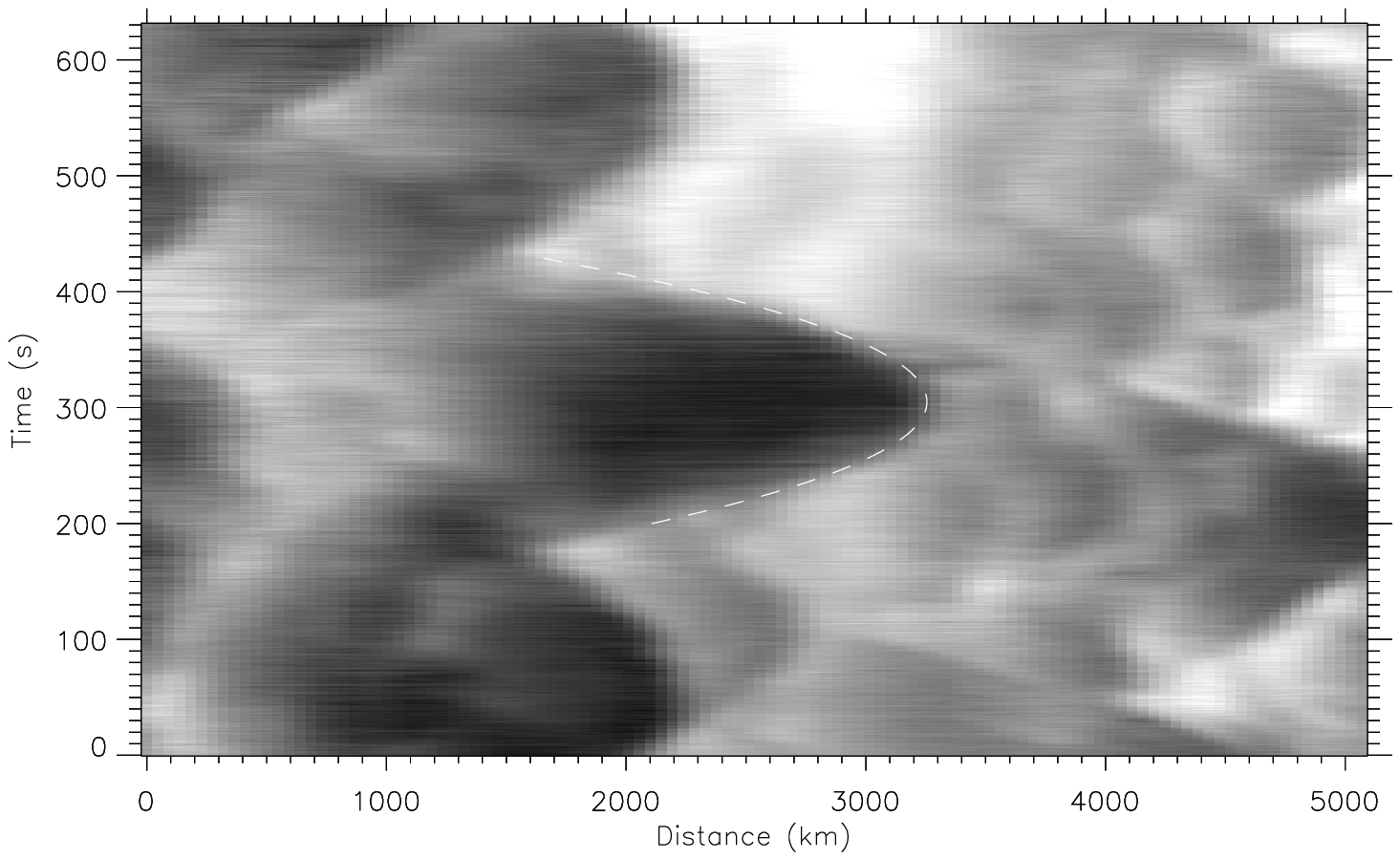}
\epsscale{0.5}
\plotone{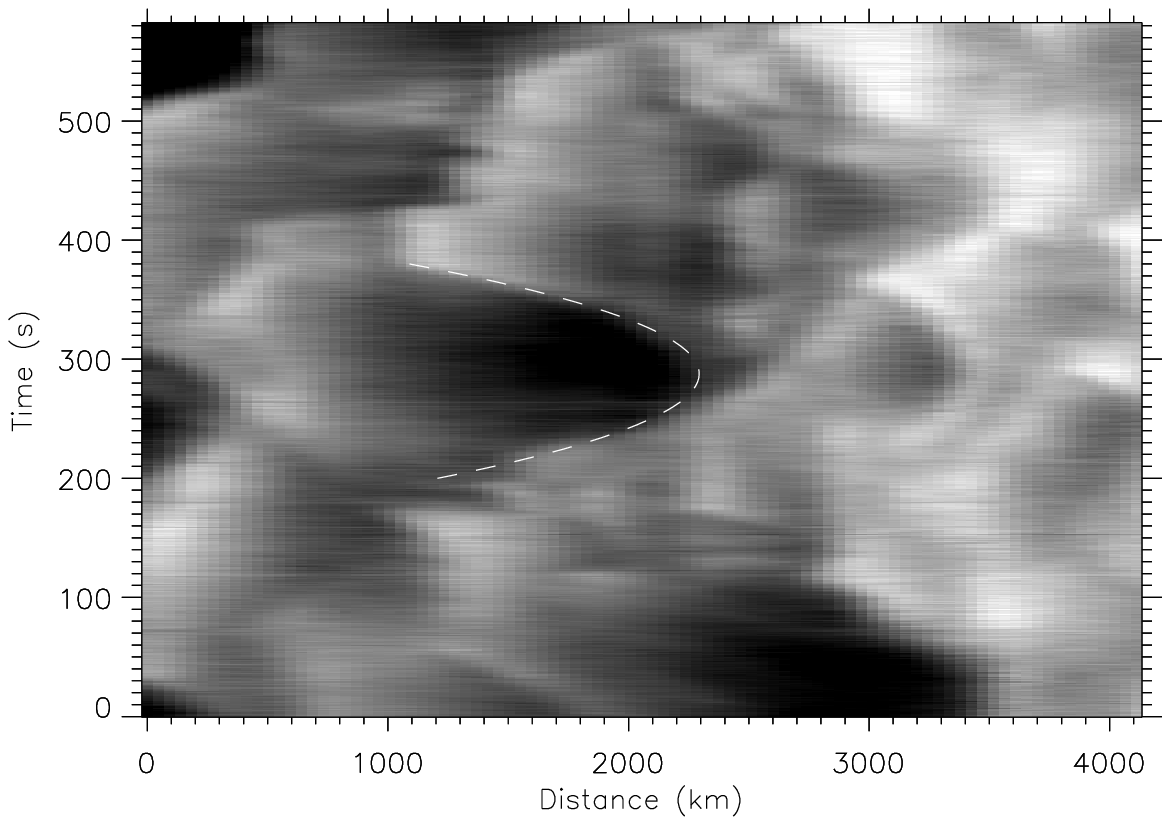}
\plotone{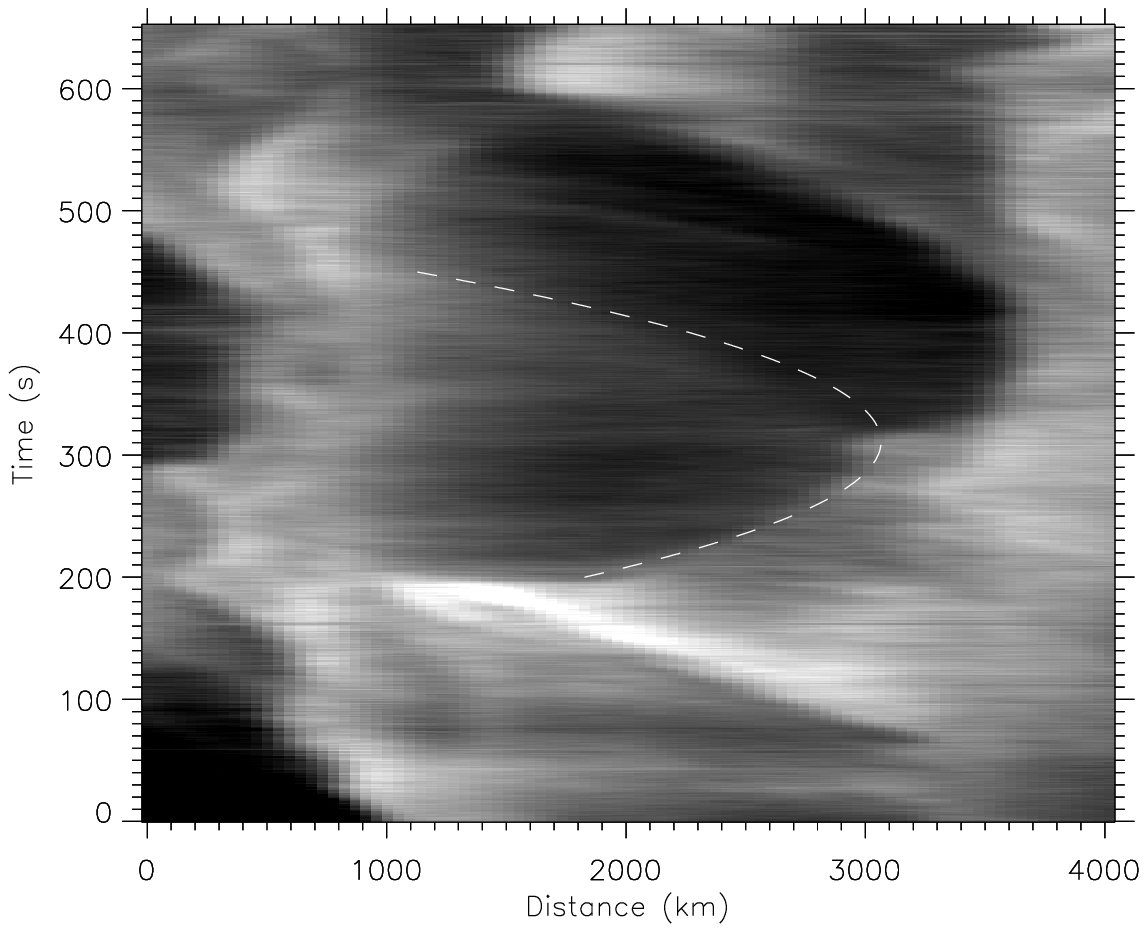}
\caption{'xt'-plots for the three DFs illustrated in
  Figs.~\ref{fig_spiclife_1}, \ref{fig_spiclife_2} and
  \ref{fig_spiclife_3}. The dynamic fibril in the top panel
  (see Fig.~\ref{fig_spiclife_1}) follows a near perfect parabolic path in
  its rise and descent. In the background
  other DFs are seen to follow similarly parabolic paths. The white
  dashed line indicates the best fit used to derive deceleration,
  maximum velocity, duration and maximum length. The dynamic fibril in
  the middle panel (Fig.~\ref{fig_spiclife_2}) also follows a near
  perfect parabolic path. It does not maintain the same brightness
  throughout its life, with slow changes occurring along the axis of
  the DF during the ascent and descent. Some minimal effects from
  changes in the atmospheric seeing quality can be seen as brighter
  stripes in the horizontal direction. The dynamic fibril in the
  bottom panel (Fig.~\ref{fig_spiclife_3}) is accompanied by a
  taller companion which seems to occur just behind it. The DF in the
  front (white dashed line) is somewhat brighter than
  the dark DF in the background.
  \label{fig_spicxt}}
\end{figure}

The length of the DF as a function of time is well described by a
parabola, as shown in the 'xt'-plot in the top panel of
Fig.~\ref{fig_spicxt}.  This is the case for the vast majority of the
DFs that do not suffer from superposition and where the DF guideline
was chosen with a correct angle, i.e. where the DF motion runs along
the guideline. All 257 DFs used in the analysis below meet these
conditions.  Each individual DF is therefore fitted with a parabola
and the deceleration, maximum velocity (on ascent or descent), maximum
extent and duration were derived. To determine the beginning and end
points of the parabola, we use the following method. The steep
parabolic ascent and descent of a typical DF (and the very high
cadence of our data) allows for a robust determination of the
beginning and end time of each DF.  The intersection (in the
'xt'-plot) between the parabolic path and a horizontal line at the
beginning time or end time is used to determine the location of the
"root" of the DF at, respectively, the beginning and end of its
lifetime. We then use these points in the "xt"-plot to fit a parabola
to the path followed by the DF.

The prevalence of parabolic paths is illustrated in all three panels
of Fig.~\ref{fig_spicxt}, where a wide variety of DFs are evident. The
DF studied in this case (top panel) follows its fitted parabola very
well with a constant deceleration given by 216~\accu, about
two thirds of solar gravity (274~\accu), a maximum velocity of 27~\kms,
and a maximum extent of some 1800~km. The properties for this DF and
all others discussed in this paper are measured along the projected
path.

An example of a much wider DF in region 2 is shown in
Fig.~\ref{fig_spiclife_2}.  This DF has extensive internal structure
and rises over a front of about $1.5$\arcsec, a width that is largely
maintained throughout the descending phase of its relatively short
lifetime of 188~s. While rising, the DF width is not constant along
its axis but the DF is wider at the bottom. It is typical that such
wide DFs do not rise and fall as a rigid body but rather show phase
and amplitude variations in velocity at various positions away from
the axis. It is also worth noting that this DF clearly becomes darker
as it rises and brightens somewhat as it falls back along the same
path it rose along. The 'xt'-plot in the middle panel of
Fig.~\ref{fig_spicxt} reveals that this DF also follows a path that is
very close to a parabola with a deceleration of 286~\accu, a maximum
velocity of 27~\kms, and a maximum extent of 1300~km.

A longer lived DF from region 1 with a lifetime of some 290~s is shown
in Fig.~\ref{fig_spiclife_3}. This DF is located in a region where
\halpha\ images indicate that highly inclined magnetic field lines
connect to the region of opposite polarity plage and pores to the
west, in the top right corner of Fig.~\ref{fig_context}. The DF is
wide, roughly 1\arcsec, and relatively long. It is also superposed on
another DF that is evident in the bottom panel of
Fig.~\ref{fig_spicxt}.  Detailed examination of the movie
  accompanying Fig.~\ref{fig_spiclife_3} shows the separate
  evolution of both DFs. The superposition makes it difficult to
determine whether these DFs are formed at the exact same location.
Residual effects of image distortion from atmospheric seeing can cause
the jagged appearance of these 'xt' profiles. Despite such distortions
and the effects of superposition, a good fit for this DF can still be
obtained with a deceleration of 200~\accu, a maximum velocity of
28~\kms, and a maximum extent of some 2000~km.

\subsection{Measurements}

The properties of the parabolic fits to the 'xt'-plots of 257 DFs were
calculated.  Visual inspection of DF time lapse movies confirmed the
proper alignment of the DFs and quality of the parabolic fits.

The upper left panel of Fig.~\ref{fig_hist} shows that the DFs suffer
a deceleration of on average 146~\accu\ with a significant spread
ranging from 40 to 320~\accu\, and a standard deviation of 56~\accu.
This is significantly less than solar gravity. Note that the median of
all DF properties is illustrated and described in the caption of
Fig.~\ref{fig_hist}, whereas we enumerate average values in the
text. Also note that these deceleration values are twice as high as
the values reported in \citet{Hansteen+etal2006}. The latter were mistakenly
divided by two.
The maximum DF lengths are on average $1\,250$~km, ranging from 400 to
$5\,200$~km with a standard deviation of 620~km, as shown in the upper
right panel of Fig.~\ref{fig_hist}. The lengths found for DFs are
smaller than the widely reported values of $5\,000$ to $10\,000$~km
for spicules at the limb \citep{Beckers1968}, but fit well with the
values reported in \citet{DePontieu+etal2004}. On average, the maximum
velocity (on either ascent or descent) is 18~\kms\, ranging from 8 to
35~\kms\ with a standard deviation of 6~\kms, as shown in the middle
left panel of Fig.~\ref{fig_hist}. The maximum velocity is never
observed to be lower than 8 \kms. DF durations are shown in the middle
right panel of Fig.~\ref{fig_hist}.  The durations vary from 120 to
650~s, with an average of 290~s, and a standard deviation of 85~s. The
median duration is lower at 250~s, because of the significant tail of
longer duration DFs.

In addition to the parabolic fits, we have also measured DF widths. To
determine the width of a DF, we calculate 'xt'-plots (with x along the
direction of the guideline) for a range of locations that are each
offset by one pixel in the direction perpendicular to the guideline.
We then find the central axis of the DF through visual inspection of
the 'xt'-plots: the location with the best-defined 'xt'-plot is chosen
as the central axis. Starting at this location we find the edges of
the DF in the direction perpendicular to the DF axis.  We search for
the two positions of maximum intensity gradient furthest from the axis
where the DF is still visible in the 'xt'-plots. This is done for two
different locations along the DF, at 50\% and at 80\% of the maximum
extent for a specific time. The width is measured in this fashion at
three different times in the DF lifetime, at 25\%, 50\% (maximum
extent), and 75\% of the DF duration. Although there is a large
spread, we find that the widths of DFs do not vary much with height
nor time. The lower left panel of Fig.~\ref{fig_hist} includes all the
widths measured and shows a mean of 340~km with a standard deviation
of 160~km. While the widest DF was measured to be 1100~km, most DFs
have widths between the difraction limit of 120~km and 380~km, with a
significant tail of wider DFs extending to 700~km.  Based on the shape
of the histogram it seems likely that we are not resolving all DFs.
However, a significant fraction is clearly resolved.
  
Visual inspection shows that the tops of most DFs are very sharply
delineated (see, e.g., Fig.~\ref{fig_spicxt}).  Furthermore,
inspection of 'xt'-plots shows that DFs often are outlined by a bright
ring, as shown in Fig.~\ref{fig_bright_ring}.  To determine how sharp
the transition at the top of the DF is, we calculated the distance
along the DF axis between the location of maximum intensity gradient
and the location of maximum intensity in the bright ring above the DF.
We refer to this measure as the thickness of the bright ring. It is
unclear what causes the bright ring. Perhaps it is related to the
steep transition from chromospheric to coronal temperatures?  Detailed
radiative transfer calculations will be necessary to determine what
causes the bright ring.  Regardless of the uncertainty in
interpretation, we measured the thickness of the bright ring for all
257 DFs and found an average thickness of 240~km with a standard
deviation of 120~km. Over half are less than 200~km and many are
unresolved, as shown in the lower right panel of Fig.~\ref{fig_hist}.

\subsection{Regional differences}

Despite a large spread in properties we find clear differences between
DF statistics in the two regions defined in the right panel of
Fig.~\ref{fig_potext}.  On average, DFs in region 1 show lower
decelerations, greater lengths, slightly higher maximum velocities,
and longer durations than in region 2 (Fig.~\ref{fig_hist}). We also
found a slight tendency for DFs to be wider in region 2 (lower left
panel of Fig.~\ref{fig_hist}), but found no difference in transition
region thickness between the two regions. Visual inspection shows that
region 2 does show more DFs that are wide with significant
substructure, such as the DF illustrated in Fig.~\ref{fig_spiclife_2}.

To further illustrate the difference between region 1 and 2, maps of
the location of DFs, color coded to show various properties, are shown
in Fig.~\ref{fig_regionalmaps}. As can be seen, there are clear
regional differences in deceleration, duration, and length.  There is
a prevalence of smaller decelerations in region 1 with typical values
around 100~\accu, versus roughly 200~\accu\ in region 2. Even more
clearly, the durations in region 1 are on average about 5~minutes,
whereas we find 3~minutes in region 2. Similarly, DF lengths in region
1 are typically around $2\,000$~km, whereas shorter DFs with lengths
around $1\,000$~km dominate region 2. The spatial pattern of maximum
velocities is not as clear with a large spread of velocities in both
regions.

In principle some of these differences could be caused by projection
effects due to the angle between the DF axis and the line of sight
vector. A DF that propagates along a direction that is parallel to the
direction of the line of sight would not be observed as a moving
feature in these \halpha\ movies. To measure the absolute extent,
velocity and deceleration along the path of the DF, the line of sight
needs to be perpendicular to the axis of the DF.  Note that the DF
durations are independent of the line of sight as they do not suffer
from these projection effects. 

To exclude the possibility that the regional differences in
deceleration, velocity and length are caused by projection effects, we
try to estimate the absolute direction of propagation of each DF.
Assuming that DFs propagate along the magnetic field, their direction
can be estimated from the potential field extrapolation shown in the
left panel of Fig.~\ref{fig_potext}\footnote{See {\tt
    http://zorak.lmsal.com/bdp/projects/vrml/spic.vrml} for a VRML
  representation. The straight red line in the VRML link is parallel
  to the line of sight.}. To estimate the effect of projection, we
calculate the angle between the line of sight vector and the direction
of the magnetic field at a reference height of $3\,000$~km above the
photosphere. This height was chosen to be very roughly the sum of the
top height of the quiescent chromosphere and the typical DF length.
The angle between extrapolated field lines and the line of sight
vector was used to correct the properties of each DF for projection.

\begin{figure}
\epsscale{1}
\plotone{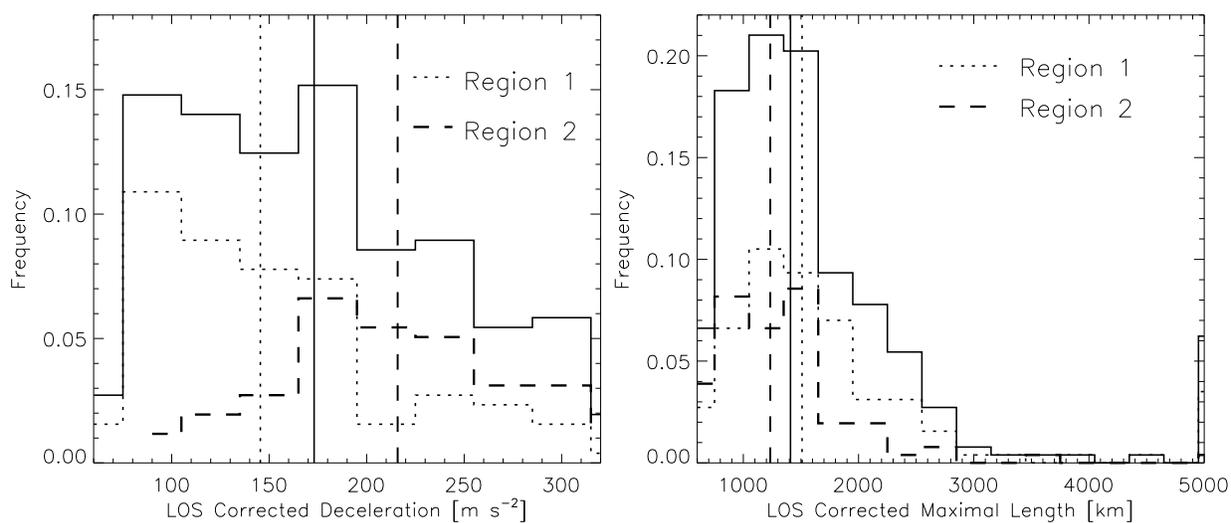}
\caption{Normalized histograms for decelerations (left) and maximum
  lengths (right) that have been corrected for line-of-sight
  projection. We attempt to calculate the deceleration and maximum
  length that would be observed when the line-of-sight is
  perpendicular to the axis of the DF. We assume that the DF axis is
  aligned with the local magnetic field, as deduced from a potential
  field calculation, and take into account the actual line-of-sight
  vector of these observations. The regional differences do not
  qualititatively change: DFs in region 1 continue to be longer and
  have lower decelerations than those in region 2.
  \label{fig_hist_los}}
\end{figure}

Histograms for these corrected values are shown in
Fig.~\ref{fig_hist_los}. Not surprisingly, all corrected values are
larger than the observed values. In addition, the corrected values
show the same spatial differences between region 1 and 2 as the
uncorrected values. To estimate how reliable the potential field
extrapolation is, we measured the mismatch between the direction of
the DF axis and the direction of the magnetic field as projected onto
the surface, and found it to be of order 20 -- 30 degrees. A mismatch
of this magnitude can lead to significant errors in calculating the
projection angle if the DF axis is nearly parallel to the projection
of the line of sight vector onto the surface. This means that the
corrected values should be considered with caution, especially since
the calculations are also based on various assumptions whose validity
is unclear. The poor resolution of the MDI magnetogram, the possible
non-potentiality of the active region, and the uncertainty in the
actual DF height with respect to the chosen reference height, all
could lead to significant errors in the estimated projection angle.
For example, the potential field extrapolation clearly fails in the
top part of region 1 where nearly horizontal fields are predicted.
Such horizontal fields are imcompatible with the visual appearance of
\halpha\ linecenter images. In addition, since horizontal fields at
this location are almost parallel to the line of sight vector, they
lead to very large and obviously erroneous corrections of the local DF
parameters.
 
Despite these reservations, it is very difficult to conceive of a
magnetic topology with an orientation to the line of sight that would
remove the regional differences for all derived parameters. In
addition, the regional differences in duration (which are independent
of projection effects) as well as the correlations found between
various parameters in the following subsection, strongly suggest that
regional differences are not caused by projection effects.


\subsection{Correlations}

Scatterplots of various DF properties reveal intriguing correlations.
The most striking correlation, also presented in
\citet{Hansteen+etal2006}, is that between the deceleration and the
maximum velocity of the DFs, in which a linear relationship is evident
(upper left panel of Fig.~\ref{fig_scattergood}). The larger the
deceleration the DF suffers from, the higher the maximum (initial or
final) velocity it shows.  Note that there are, again, significant
differences between region 1 and 2: the DFs in region 2 typically have
larger decelerations for a given maximum velocity. A linear fit to the
observed relationship between the deceleration and velocity shows that
both regions have almost identical slopes.

Another interesting correlation is that between the maximum length and
the duration of a DF (upper right panel of
Fig.~\ref{fig_scattergood}). The longer DFs tend to have longer
lifetimes. As expected from the histograms in the upper and middle
right panels of Fig.~\ref{fig_hist}, the DFs in region 1 gravitate
towards the upper right of this scatterplot. We also find that the
maximum velocity and maximum length of DFs is well correlated (lower
right panel of Fig.~\ref{fig_scattergood}): DFs with higher maximum
velocity tend to be longer. However, there is a difference in the
slope in regions 1 and 2. A linear fit for region 1 reveals a slope of
87 km per \kms, and 60 km per \kms\ for region 2. DFs in region 1 are
longer for the same maximum velocity.

The deceleration of DFs shows a somewhat less clear correlation with
the DF duration (lower left panel of Fig.~\ref{fig_scattergood}). The
longest-lived DFs typically suffer from the lowest deceleration. This
correlation shows a large spread, and is not quite linear for the
total population of DFs. It seems that each region may have its own
linear relationship. DFs in region 1 show a more steep dependence of
duration on deceleration, whereas those in region 2 show a smaller
range of durations, for a range of decelerations that is similar in
extent to that of region 1.

Based on the correlations described in the above, one would perhaps
also expect a correlation between the deceleration and maximum length,
or between the maximum velocity and duration. However, such
correlations are not clear from the observations, as demonstrated in
Fig.~\ref{fig_scatterbad}. The large scatter and different slopes that
are evident in the plots that do show correlations may be partially
responsible for the lack of clear correlation between deceleration and
maximum length or the maximum velocity and duration. However, it is
also clear that there are very significant regional differences in
Fig.~\ref{fig_scatterbad}, with each region occupying a different area
in the scatterplots, and in some cases showing different slopes. It
seems probable that such regional differences also are responsible for
the lack of clear correlations.

When we take the line-of-sight correction into account, none of these
correlations, or the lack of correlation, change qualitatively.
Evidently, the correlation between deceleration and velocity, or that
between maximum velocity and length cannot change significantly
because of projection effects, since both parameters would be
corrected by the same factor. The fact that the other correlations,
between duration on the one hand, and either deceleration, length or
velocity, do not change qualitatively indicates that projection
effects do not dominate our dataset. 

The same arguments also apply to any kind of selection effects that
are caused by the fact that DFs with very large line-of-sight
velocities could disappear from or be invisible in our relatively
narrow (128 m\AA) \halpha\ linecenter bandpass. Since DFs are much
wider spectral features 
\citep[with widths of order 1000 m\AA, see, e.g.][]{Beckers1968} 
than the width
of the bandpass, this means that only features with line-of-sight
velocities of order 20 or more km/s would be invisible or only partially 
visible in the linecenter timeseries. While in a few cases we can
see some fading of the DF during its descent, it is still possible to
trace the top of the DF to its end point. Given these arguments, the
passband effect can only change the derived correlations marginally.

Note that all of the measurements on which the correlations are based,
depend on the identification of a single DF rising above the
chromospheric background. The latter can be quite noisy and sometimes
less well defined. Filtergrams taken in the wings of H$\alpha$ might
be helpful for a more exact determination of the root location of
each DF, but such filtergrams were unavailable in this dataset.  This
means that the measured length and maximum velocity could in fact be
somewhat higher than reported here. However, given the rapid rise and
fall of DFs, the measured duration is quite insensitive to this
effect. Since we are measuring the acceleration by fitting parabolas
to the height profile, the acceleration is also insensitive to the
position of the chromospheric background. Overall, this implies that
the relative uncertainty in determining the roots of DFs are unlikely
to qualitatively change the correlations reported here.

The interpretation of these correlations, or the lack thereof, will be
discussed at length in \S \ref{section:discussion}.


\subsection{Oscillations}

The regional differences in DF durations evident in the middle right
panel of Fig.~\ref{fig_hist}, with 3 minute lifetimes dominating
region 2 and lifetimes around 5 minutes more dominant in region 1,
suggest a role for chromospheric waves or oscillations in the
formation mechanism of DFs. Oscillations in the chromosphere are
typically not steady harmonic waves, but wave trains of finite
duration. For this reason we perform a wavelet analysis of the
\halpha\ data, since it allows for localization in time of periodic
signals.  The mother wavelet we use is the complex valued Morlet
wavelet ($k=6$) which consists of a plane wave modulated by a
Gaussian, an appropriate shape for wave trains. To estimate the
statistical significance of the wavelet power spectra (95 \%
confidence interval), we compare them to theoretical spectra for white
noise, following \citet{Torrence+Compo1998} and
\citet{DeMoortel+Hood2000}. Taking into account the cone of influence
\citep{Torrence+Compo1998}, we calculate for each superpixel the number of
wavepackets in the \halpha\ timeseries for which the wavelet power is
significant. This is done for a range of wave periods from 150 to 600
seconds. A superpixel is defined as the average of 4x4 original
0.063\arcsec\ pixels. Note that the original data is oversampled by a
factor of almost 3 with respect to the diffraction limit (120 km) of
the SST at the wavelength of \halpha.

\begin{figure}
\epsscale{0.9}
\plotone{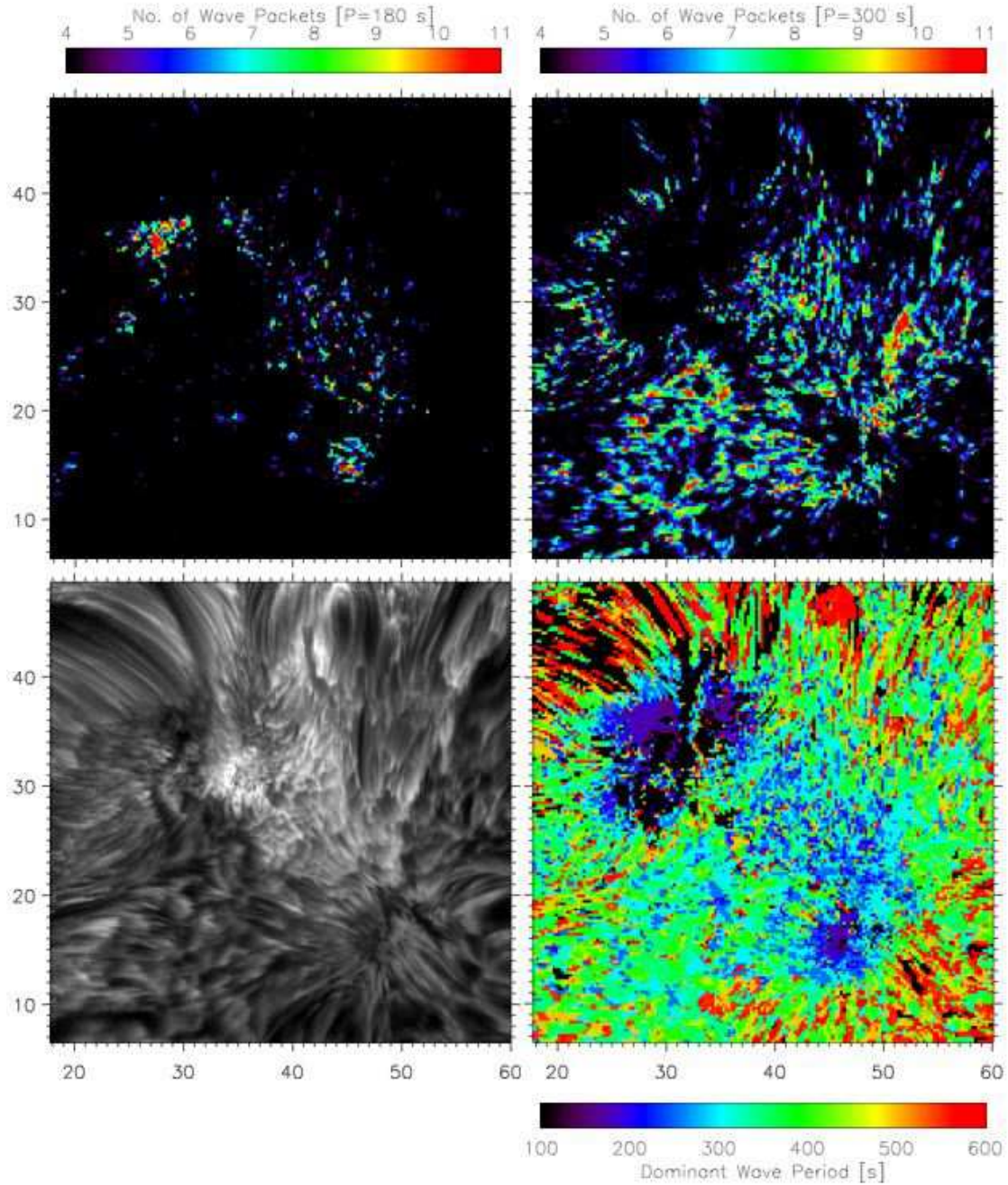}
\caption{Results from a wavelet analysis of the 78 minute long
  \halpha\ linecenter timeseries, a snapshot of which is shown in the
  lower left panel. The two top panels show, for each location of the
  lower left panel, the number of significant wave packets detected
  for waves with periods of 180 s (left) and 300 s (right). Locations
  with less than 4 wavepackets are not considered locations with
  significant power (shown as black). The lower right panel
  illustrates for each location which wave period dominates, i.e.,
  contains the highest number of wavepackets with significant power.
  The two sunspots and region 2 are dominated by 3 minute power,
  region 1 by 5 minute power, and the highly inclined fibrils (e.g.,
  in the superpenumbra) by oscillations with even longer periods.
  \label{fig_osc_panels}}
\end{figure}

The upper left and upper right panels of Fig.~\ref{fig_osc_panels}
show, for each location of the lower left panel, the number of
significant wavepackets (during the \halpha\ timeseries) for waves
with a period of 180 s and 300 s respectively.  The lower right panel
shows which wave period dominates, i.e., has the most wavepackets, at
each location of the field of view. From this figure, we find that
many locations show significant oscillatory power throughout the
timeseries. The dominant period of this power varies across the field
of view. Clearly visible are the dominant 3 minute oscillations in the
sunspots around x=27\arcsec\ and y=37\arcsec\ and at x=45\arcsec\ and
y=15\arcsec. In addition, the dense plage region which we have
designated as region 2, shows a preponderance of 3 minute
oscillations. In contrast, region 1, adjacent to the dense plage
region, is dominated by oscillations with long periods around and
above 5 minutes. In this region, the magnetic field is more inclined
than in the dense plage region. The longer, less dynamic and more
horizontal fibrils, such as those in the superpenumbra of both spots,
seem to be dominated by oscillations with periods longer than 10
minutes.

It is interesting to note that the dominant period in region 2
corresponds exactly to the DF duration we observe in region 2.
Apparently, DFs typically live for about 3 minutes in this region,
which is also the dominant periodicity of the oscillations. Likewise,
region 1 is dominated by 5 minute oscillations, and contains DFs with
lifetimes of order 5 minutes. Since the oscillation measurements are
based on the same data, it seems clear that some DFs are periodic or
recurring. This was also described by \citet{Hansteen+etal2006} and
suggests a relationship between chromospheric waves and DFs, which
will be further explored in \S \ref{section:discussion}.

\section{Discussion}
\label{section:discussion}

\subsection{Observational Findings}

\begin{figure}
\epsscale{0.7}
\plotone{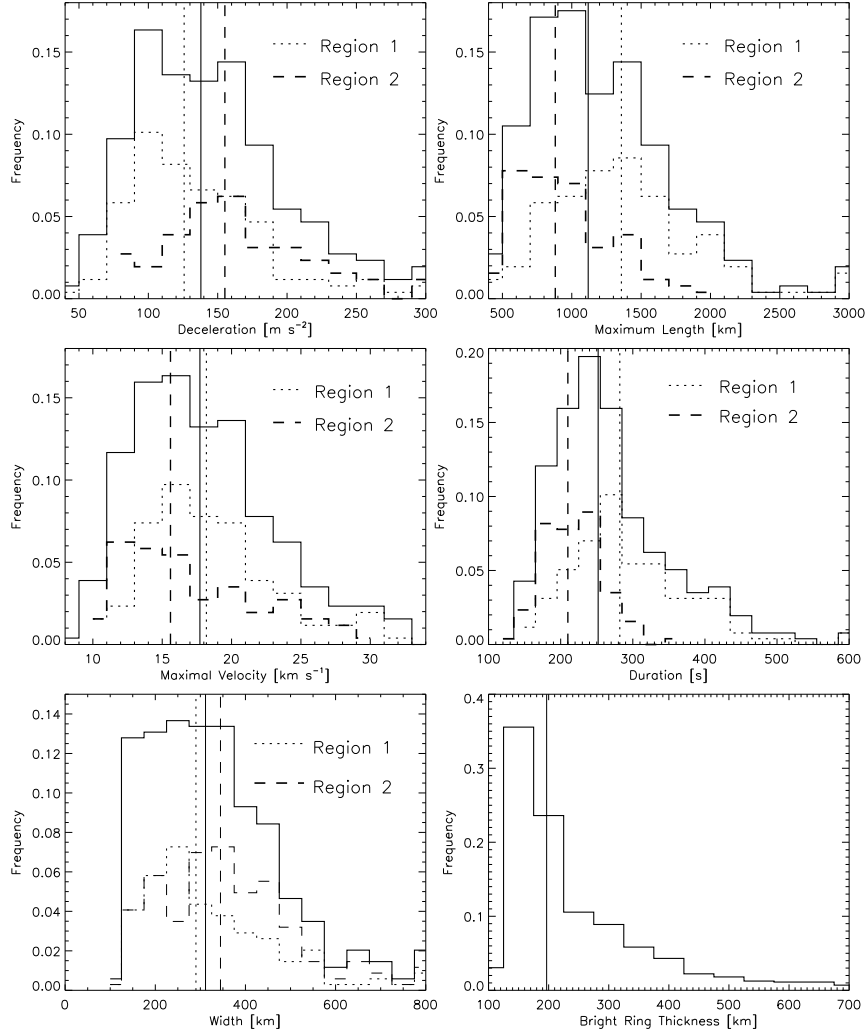}
\caption{Normalized histograms for decelerations (upper left), maximum
  lengths (upper right), maximum velocity (middle left), duration
  (middle right), width (lower left) and thickness of the bright ring
  (lower right), measured for all 257 DFs (except widths which were
  measured for 91 DFs). The full vertical line in all plots indicates
  the median value of the distribution: a deceleration of 136~\accu,
  maximum length of 1100~km, maximum velocity of 17.8~\kms, duration
  of 250~s, width of 310~km and bright ring thickness of 190~km.
  The two regions illustrated in the right panel of
  Fig.~\ref{fig_potext} show clearly different distributions of the
  deceleration, maximum lengths and duration, as well as some
  differences in maximum velocity and width. Region 1 is indicated by
  dotted histograms and dotted vertical lines for the median values.
  This region shows lower decelerations (126~\accu), longer durations
  (280~s) and lengths (1350~km), slightly larger velocities
  (18.2~\kms) and slightly smaller widths (290~km) than region 2.
  Region 2 is indicated by dashed histograms and dashed vertical lines
  for the median values. It has higher decelerations (150~\accu),
  shorter durations (210~s) and lengths (880~km), slightly lower
  velocities (15.4~\kms) and sightly larger widths (340~km). The
  regional histograms are normalized to the total number of DFs.
  \label{fig_hist}}
\end{figure}

\begin{figure}
\epsscale{0.5}
\plotone{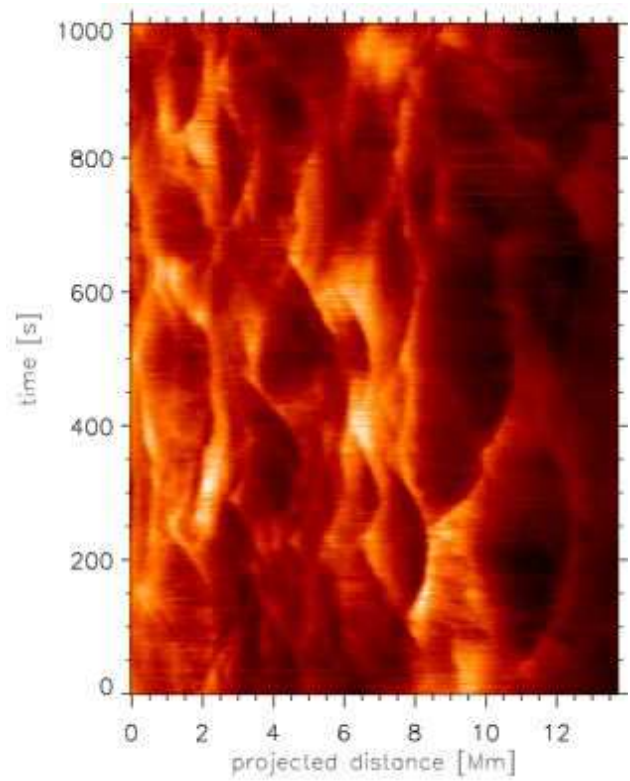}
\caption{'xt'-plot for several DFs showing bright rings at their top
  end. This bright emission could be a measure of how steep the
  transition to hotter temperatures is at the top of the DF. Most DFs
  show some maximum of brightness at their top end. Measurements of
  the gradient at the top of DFs are given in the lower right panel of
  Fig.~\ref{fig_hist}.
  \label{fig_bright_ring}}
\end{figure}

Most DFs follow parabolic paths with a symmetrical ascending and
descending phase. These paths are characterized by a large initial
velocity, usually of order 15-20 \kms\ (middle left panel of
Fig.~\ref{fig_hist}), that decreases linearly with time until the DF
has retreated completely. At the end of its life, a DF's downward
velocity is roughly identical to the initial upward velocity. The
linear decrease with time of the velocity, i.e., the deceleration, is
usually between 120 and 280 \accu\ (upper left panel of
Fig.~\ref{fig_hist}).  There is a clear linear relationship between
the deceleration and the maximum (down- or upward) velocity of the DF
(upper left panel of Fig.~\ref{fig_scattergood}).

DFs reach maximum lengths of order 1,000 to 2,000 km (upper right
panel of Fig.~\ref{fig_hist}) during their lifetimes of 3 to 8 minutes
(middle right panel of Fig.~\ref{fig_hist}). The duration and maximum
length are well correlated (upper right panel of
Fig.~\ref{fig_scattergood}), which indicates that uncertainty about
the line-of-sight correction may not play an important role in the
regional differences we find. We resolve more than half of the DFs we
observe. The average thickness is 240 km (lower left panel of
Fig.~\ref{fig_hist}), with a range between the diffraction limit (120
km) and 1000 km. Some DFs occur over a wider spatial area, whereas
others are very thin and stay thin throughout their lifetime. Many of
the wider DFs have variable widths as a function of time. These give
the visual appearance of a wave front with varying phase speeds or
amplitude in the direction perpendicular to the DF axis.  There
is often a brightening above the DF, which can be clearly seen in
'xt'-plots (Fig.~\ref{fig_bright_ring}). This brightening might be a
sign of increased temperatures at the top of the DF. The thickness of
this transition is not resolved and is less than 120 km in more than
half of the DFs studied (lower right panel of Fig.~\ref{fig_hist}).

\begin{figure}
\epsscale{1}
\plotone{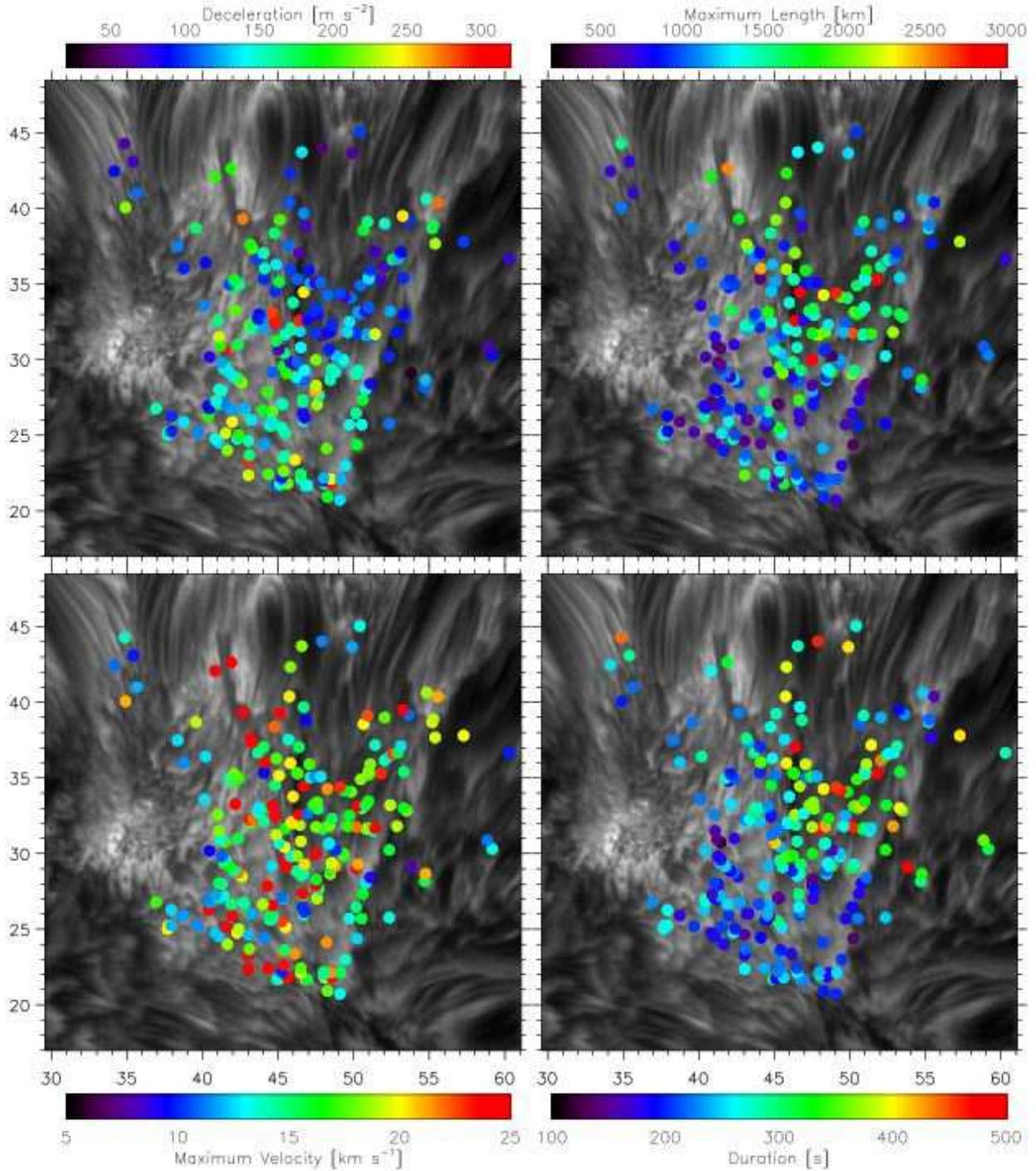}
\caption{Differences in deceleration (upper left), maximum length
  (upper right), maximum velocity (lower left) and duration (lower
  right) for all measured DFs are illustrated as color-coded dots
  superposed on a \halpha\ linecenter image. DFs in region 1 (see
  right panel of Fig.~\ref{fig_potext}) show lower deceleration,
  longer maximum length, slightly higher maximum velocity and longer
  duration than region 2.
  \label{fig_regionalmaps}}
\end{figure}
\begin{figure}
\epsscale{1}
\plotone{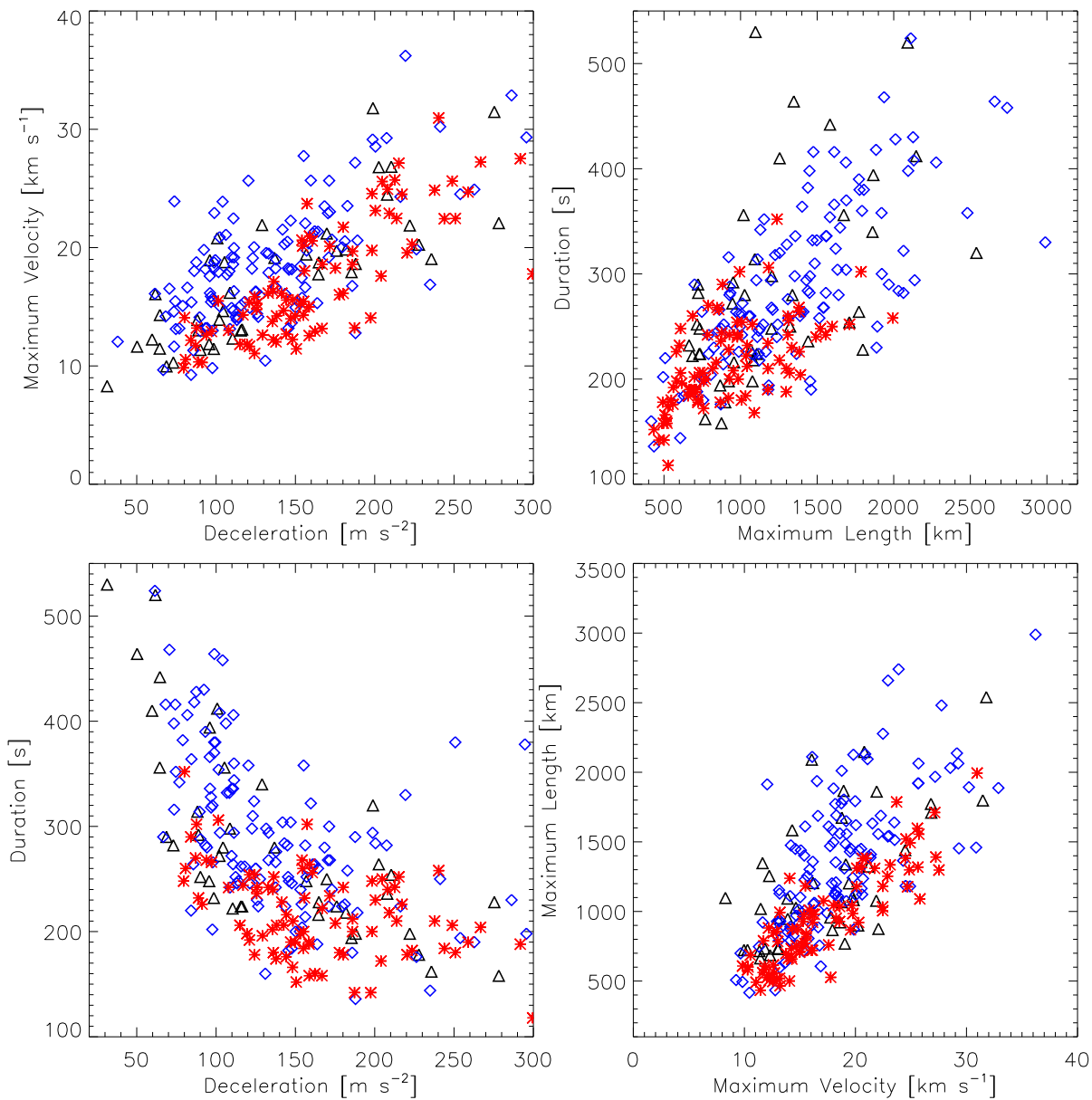}
\caption{Scatterplots of deceleration versus maximum velocity (upper left), maximum
  length versus duration (upper right), deceleration versus duration
  (lower left), and maximum velocity versus maximum length (lower
  right). Clear linear correlations exist between deceleration and
  maximum velocity, maximum length and duration, and maximum velocity
  and maximum length. There is also a less well defined anti-correlation
  between deceleration and duration. DFs with higher maximum velocity
  typically suffer more deceleration (upper left), yet reach greater
  maximum lengths (lower right). DFs that are longer typically last
  longer. There are clear regional differences between region 1 (blue
  diamonds) and region 2 (red stars), which are also illustrated in
  Figs.~\ref{fig_hist} and \ref{fig_regionalmaps}. The
  scatterplots for the two regions often seem to have different
  slopes.
  \label{fig_scattergood}}
\end{figure}
\begin{figure}
\epsscale{1}
\plotone{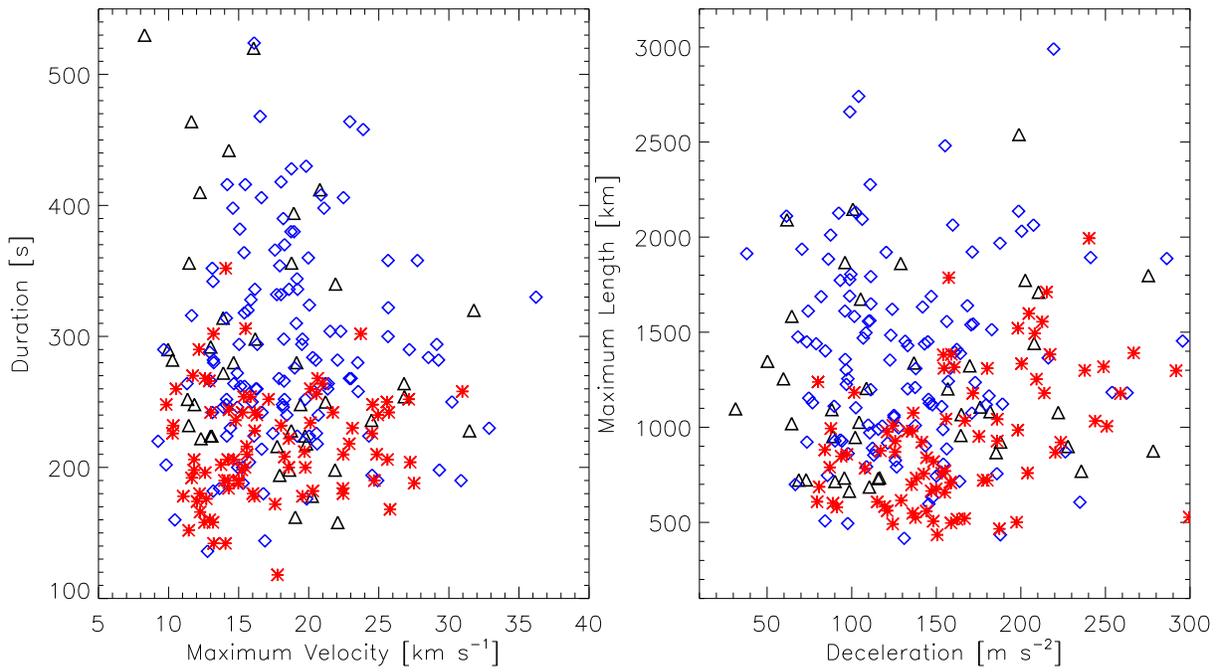}
\caption{Scatterplots of maximum velocity versus duration, and
  deceleration versus maximum length. These plots do not show clear
  correlations. This may be caused by the wide spread in the
  scatterplots of Fig.~\ref{fig_scattergood}, and the regional
  differences in many of the correlations. 
  \label{fig_scatterbad}}
\end{figure}

DFs have significantly different properties in two regions within our
field of view. Region 2 contains dense plage, i.e., a collection of
strong magnetic field concentrations in which the magnetic field is
generally more vertical. Region 1 is located at the edge of this dense
plage region, where the magnetic field is more inclined, as it
connects to relatively nearby opposite polarity plage. The DFs in the
dense plage region are shorter ($\sim1,000$ km), have higher
decelerations ($\sim200$ \accu), slightly lower velocities ($\sim 15$
\kms) and shorter durations of about 3 minutes (Figs.~\ref{fig_hist}
and \ref{fig_regionalmaps}). In contrast, the DFs in the region where
the field is more inclined are longer ($\sim 2,000$ km), have lower
decelerations ($\sim 120$ \accu), slightly higher velocities ($\sim 20$
\kms) and longer durations of about 5 minutes (Figs.~\ref{fig_hist}
and \ref{fig_regionalmaps}). In addition, DF lifetimes correlate well
with the dominant periodicities observed in both regions: 3 minute
oscillations dominate \halpha\ in the dense plage region, whereas 5
minute oscillations dominate in the region with more inclined field
(Fig.~\ref{fig_osc_panels}).

\subsection{Interpretation}

Our measurements of DFs and oscillatory power strongly suggest that
chromospheric waves drive DFs upward. These chromospheric waves are
generated by convective flows and (global) oscillations in the
photosphere and convection zone. The magnetic field concentrations
provide a channel for these disturbances to propagate up through the
chromosphere. During their path upwards, they shock and drive
chromospheric plasma upwards, forming DFs in the process. This
mechanism was originally suggested in rudimentary form by
\citet{Parker1964}. Some models have further studied the effects of
the convective part of the photospheric driver on the chromosphere in
the so-called rebound-shock model
\citep{Hollweg1982,Sterling+Hollweg1989,Sterling+Mariska1990}. More
recently, \citet{DePontieu+etal2004} included both the convective
driver and global oscillations in varying magnetic topologies to
directly describe their (lower resolution) observations of dynamic
fibrils. The latter model's predictions agree well with the current
findings. 

This scenario is confirmed by analysis of advanced numerical
simulations (see \citet{Hansteen+etal2006} and next subsection) in
which dynamic fibril-like features with properties identical to those
in our data occur.  From the simulations, it is clear that the
observed parabolic paths indicate that DFs are shock driven.  When a
shock hits the transition region (i.e, the top of a DF), the
transition region is catapulted upwards at a velocity that exceeds the
sound speed in the chromosphere ($\sim 10$ \kms).  Shock waves
generally have velocity profiles in the form of 'N'- or 'sawtooth'
shapes \citep[e.g.][]{Mihalas+Mihalas1984}.  Thus, a plasma parcel
passing through a shock wave will first experience a sudden kick in
velocity ---- the shock itself --- and thereafter a gradual linear
deceleration as the shock recedes. Assuming a shock wave with
amplitude or shock strength of $u_0$ and linear deceleration $a$
\[ u(t)=-at+u_0, \]
it is trivially verified that a plasma parcel passing through the shock
will describe a parabolic path; integrating gives a parcel position that
moves as
\[ \xi(t)=-{1\over 2}at^2+u_0t+\xi_0. \] 
DF formation from chromospheric shocks is compatible
with the fact that we do not find any DFs with initial (maximum)
velocities of less than 10 \kms. Waves with smaller amplitudes will
not form shocks and hence DFs. The observed lower limit to the
maximum velocities provides an estimate of the sound speed in the
upper chromosphere (10 \kms). Its value is compatible with
chromospheric temperatures from solar models. More generally, the
observed maximum velocities provide a robust estimate of the Mach
number of chromospheric shocks. Our observations indicate that shocks
with Mach numbers of 1 through at least 3.5 are ubiquitous in the
magnetized chromosphere, and drive DFs.

Analysis of numerical simulations \citep{Hansteen+etal2006} further
shows that several different effects influence the shock formation and
evolution so that the observed correlations (e.g., between
deceleration and velocity) and regional differences follow naturally.

The oscillatory power spectrum of the photospheric driver is dominated
by power at 5 minutes. The chromosphere acts as a filter to the input
from below, filtering out waves with periods that are longer than the
local acoustic cutoff period.  In general, the acoustic cutoff period
depends on the inclination of the magnetic field lines to the vertical
\citep{Suematsu1990,DePontieu+etal2004}. Under conditions of vertical
magnetic field (e.g., region 2), this means that the chromosphere is
dominated by oscillations and waves with periods at the acoustic
cutoff of 3 minutes. As a result, the DFs formed in the dense plage of
region 2 are driven by shocks with periods of 3 minutes and thus have
lifetimes of order 3 minutes. In contrast, the inclined field, such as
that found in region 1, increases the acoustic cutoff period. This
allows the photospheric waves with periods of 5 minutes to leak into
the chromosphere, develop into shocks and form DFs with similar
lifetimes.

Since most of the oscillatory power in the photosphere occurs for
periods of 5 minutes, the driver in inclined field regions is
generally stronger than in regions with more vertical field. This
effect is enhanced by the fact that, in dense plage regions, the
amplitude of convective flows and global oscillations is generally
reduced at the photospheric level. DFs that form along inclined field
lines also experience only the component of gravity along the magnetic
field, so that decelerations are generally lower in the inclined field
region.

The observed correlation between deceleration and maximum velocity can
be understood in terms of shock wave physics and the fact that Dfs are
driven by single shocks. For a given wave period, 'N'-shaped shock
waves will show a steeper decline with time in velocity for a greater
shock strength. In other words, DFs with higher maximum velocity,
i.e., those that are driven by stronger shocks, will show greater
deceleration, for a given DF lifetime.


It is the combination of these effects that explains the observed
regional differences and correlations. Region 2 contains more vertical
field, so that only 3 minute power drives the chromospheric waves and
shocks. Given the photospheric power spectrum and the reduced power in
dense plage regions, this leads to lower amplitudes, i.e., lower
maximum velocity. In addition, the short lifetimes and more vertical
field lead to higher decelerations, and shorter lengths. Conversely,
region 1 contains more inclined field, allowing the full photospheric
peak power at 5 minutes to leak into the chromosphere to drive DFs. As
a result, the velocity amplitude is greater, and the longer lifetimes
and less vertical field lead to lower decelerations and longer
lengths.

\subsection{Numerical Simulations}

These findings are confirmed when comparing the observations with
recent advanced numerical radiative MHD simulations that include the
convection zone, photosphere, chromosphere, transition region and
corona \citep{Hansteen+etal2006}. The equations of MHD are solved
using an extended version of the numerical code described in
\citet{Dorch+Nordlund1998,Mackay+Galsgaard2001} and in more detail by
Nordlund \& Galsgaard at {\tt http://www.astro.ku.dk/$\sim$kg}.  The
code functions by using a sixth order accurate scheme to determine
partial spatial derivatives. The equations are stepped forward in time
using the explicit 3rd order predictor-corrector procedure by
\citet{Hyman1979}, modified for variable time steps.  In order to
supress numerical noise, high-order artificial diffusion is added both
in the forms of a viscosity and in the form of a magnetic diffusivity.
Thermal conduction along the magnetic field in the corona and
transition region is treated by operator splitting and using a
multi-grid solver.  The radiative flux divergence from the photosphere
and lower chromosphere is obtained using the method of
\citet{Nordlund1982} based on group mean opacities as modified by
\citet{Skartlien2000} to account for scattering. For the upper
chromosphere, transition region and corona we assume effectively thin
radiative losses. For the middle and upper chromosphere we parametrize
the optically thick radiative losses in strong lines and continua from
hydrogen and singly ionized calcium by using escape probabilities
calculated in 1D with the code of
\citet{Carlsson+Stein1992,Carlsson+Stein1994,Carlsson+Stein1995,Carlsson+Stein1997,Carlsson+Stein2002}.
Further description of the 3D code may be found in
\citet{Hansteen+Gudiksen2005}.

The 2d simulations described here are run on a grid of $512\times 150$
points spanning $16\times 12$~Mm$^2$. In height, these models cover a
region from 2~Mm below to 10~Mm above the photosphere.
The grid has uneven spacing in the vertical, $z$, direction with a
grid size of 30~km in the photosphere and chromosphere becoming
gradually larger in the corona. We have also run 2d calculations on
coarser grids of $256\times 150$ and $128\times 150$ points, as well
as 3d runs of $128\times 64\times 150$ points spanning $16\times
8\times 12$~Mm$^3$, with results similar to those reported here.

\begin{figure}
  \epsscale{1} \plotone{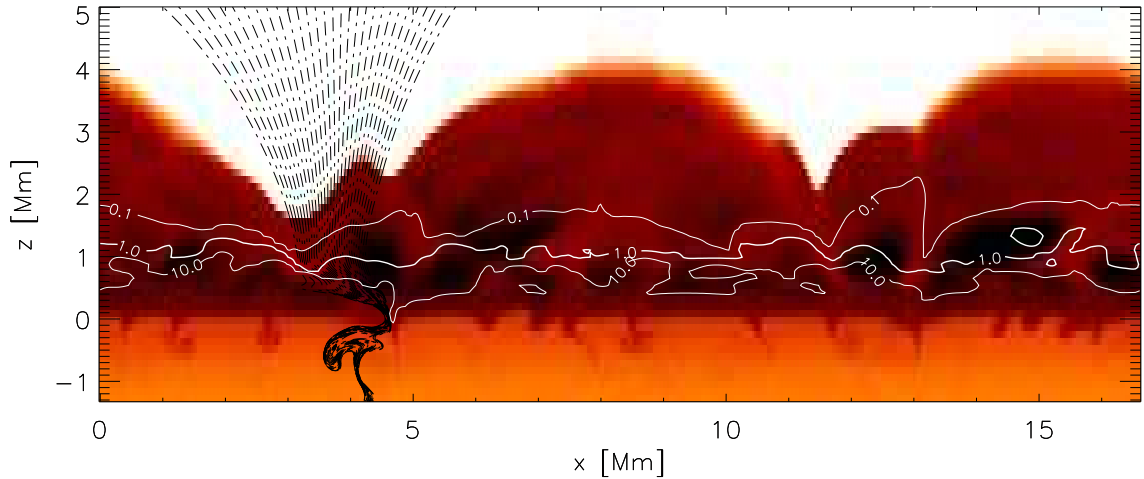}
\caption{Snapshot taken from one of the 2D numerical experiments simulating the
  generation of a dynamic fibril. The logarithm of the temperature,
  $T_{\rm g}$, is shown, set to saturate at $\log T_{\rm g}=4.5$; the
  minimum temperature is roughly 2000~K ($\log T_{\rm g}=3.3$). The
  vertical scale has its origin at the height where $\tau_{500}=1$.
  Contours of plasma $\beta$ are drawn in white where $\beta=0.1, 1,
  10$, with the $\beta=1$ contour thicker for clarity.  In black are
  drawn magnetic field lines covering the region where dynamic fibrils
  ascend as a result of upwardly propagating shock waves. We find
  events that resemble observed dynamic fibrils in this region as well
  as in the corresponding opposite polarity region centered on
  $x=12$~Mm. Note the highly intermittent nature of the chromospheric
  temperature structure and the ubiquity of shocks outlined by regions
  of high $T_g$.  These shock waves seem to preferentially enter the
  corona where the magnetic field lines also enter the corona. The
  position of the transition region does not change much in the
  regions between $x=5$~Mm and $x=12$~Mm, where the field is more
  horizontal. An mpeg-movie illustrating the dynamic evolution of the
  temperature and velocities is available online.}
  \label{sim_snapshot}
\end{figure}

At the start of a given simulation, a potential magnetic field is
added to a model of the solar atmosphere. The resultant model is then
allowed to evolve in time until a quasi-steady state is achieved. This
takes on the order of a half hour solar time. The models are
convectively unstable due the radiative losses in the photosphere.
This results in a temperature structure similar to that derived in
semi-empirical solar models. The average temperature at the bottom
boundary is maintained by setting the entropy of the fluid entering
the computational domain. The bottom boundary, based on characteristic
extrapolation, is otherwise open, allowing fluid to enter and leave
the computational domain as required. The magnetic field at the lower
boundary is advected with the fluid.  To prevent coronal cooling the
upper temperature boundary is set to evolve towards a temperature of
$800\,000$~K on a timescale of 10 -- 100~s. (In the 3d models coronal
temperatures are maintained by self consistent magnetic dissipation.)
A snapshot of the temperature structure in the lower part of the model
including the convection zone, photosphere, and chromosphere is shown
in Fig.~\ref{sim_snapshot}.

\begin{figure}
\epsscale{1}
\plotone{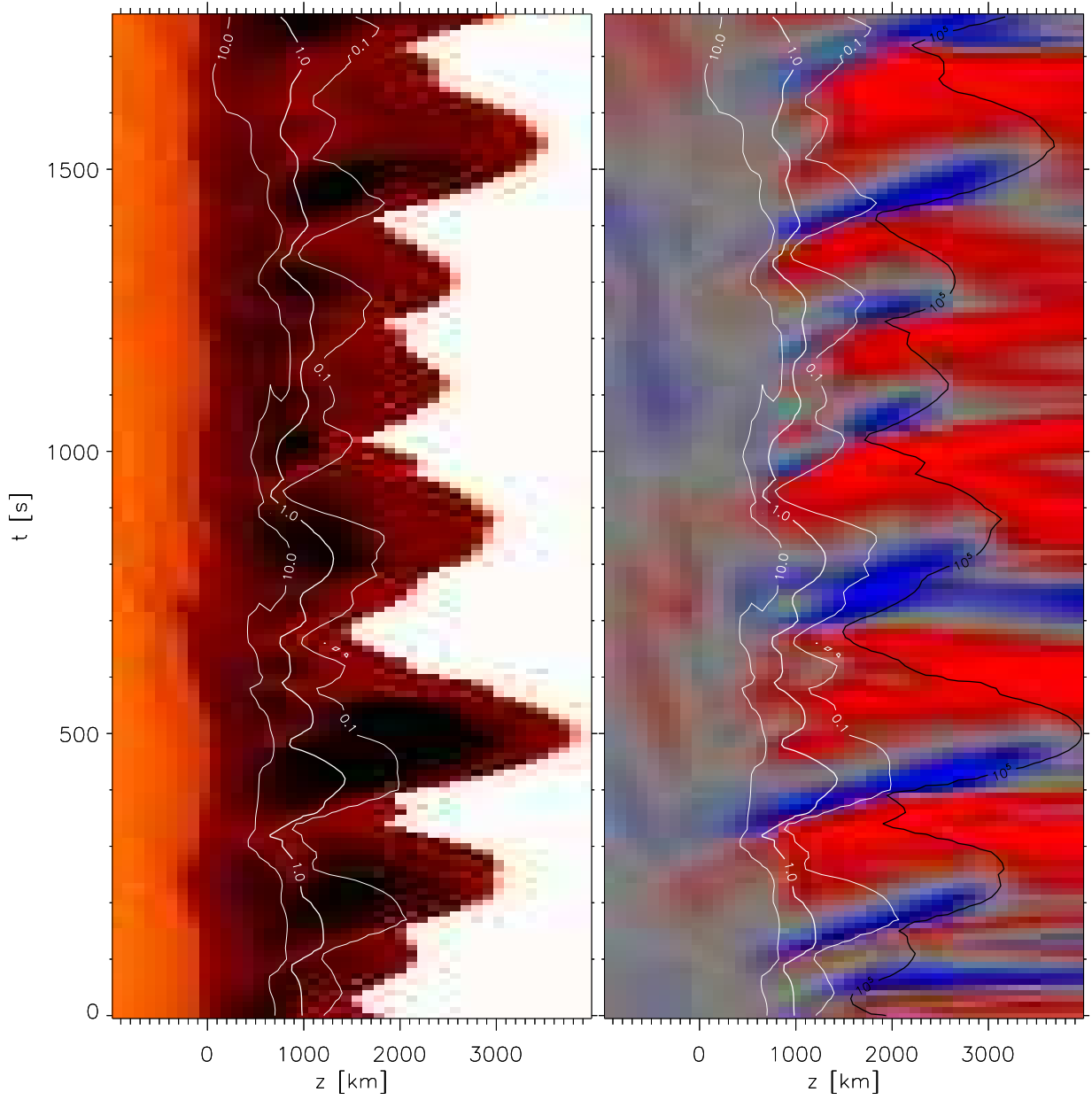}
\caption{'xt'-plot of $\log T_{\rm g}$ (left panel) and the vertical 
  velocity (right panel) as a function of height and time at horizontal 
  position $x=4$~Mm in Fig.~\ref{sim_snapshot}. The logarithm $\log
  T_{\rm g}$ of the temperature is set to saturate at $\log T_{\rm
    g}=4.5$; the minimum temperature is roughly 2000~K ($\log T_{\rm
    g}=3.3$). The velocity saturates at $\pm 20$~\kms, red color indicates
    downflows. Countours of plasma $\beta$ are drawn in white where
  $\beta=0.1, 1, 10$. A contour indicating the position of the transition
  region (where $T_{\rm g}=10^5$~K) is shown in black in the right panel.
  Note how waves, generated as linear waves in the
  photosphere and forming shocks roughly $1000$~km above
  $\tau_{500}=1$, propagate upwards and push the uppper chromosphere
  and transition region to great heights, several thousand kilometers
  above their equilibrium positions. The general appearance of this
  'xt'-plot is very similar to observed 'xt'-plots
  (Fig.~\ref{fig_spicxt}).
  \label{sim_xt}}
\end{figure}

Waves generated by convection propagate through the chromosphere,
steepen and form shocks roughly 1~Mm above the photosphere. In regions
above the level where plasma $\beta = 1$ the propagation is modified
by the magnetic field and the waves are subjected to reflection,
refraction and mode coupling as described by \citet{Bogdan+etal2003}.
Of interest to us here are the shocks that propagate along field lines
that enter the coronal plasma as indicated in Fig~\ref{sim_snapshot}.
These shocks lift and accelerate the transition region as they
propagate into the corona. Plotting the motion of the transition
region along one such field line as a function of time, e.g. a 'xt'
plot, reveals that the position of the transition region describes a
parabola (Fig.~\ref{sim_xt}, compare with Fig.~\ref{fig_spicxt}). In
the left panel of Fig.~\ref{sim_xt} the logarithm of the temperature
is shown, the right panel shows the velocity. It is worth noting that
except for a brief moment in the early ascending phase the simulated
DF shows both upflowing and downflowing velocities along its length at
a given time. During descent, the DF has only downflowing plasma. Note
also that near its base the simulated DF shows downflows for a major
portion of its lifetime.

\begin{figure}
\epsscale{0.5}
\plotone{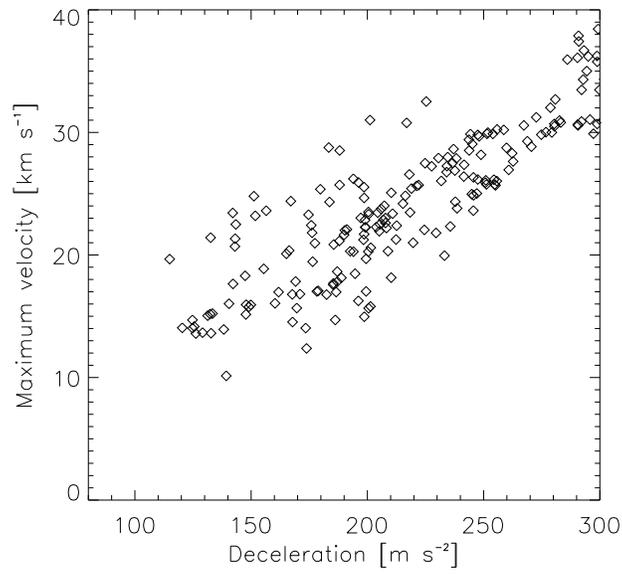}
\caption{Scatterplot of deceleration versus maximum velocity from
  measuring several DF-like features in the numerical simulations. A
  comparison with the upper left panel of Fig.~\ref{fig_scattergood},
  reveals that the simulations reproduce the observed relationship,
  and range in deceleration and maximum velocity quite well.
  \label{sim_dec-maxv}}
\end{figure}

We have measured the transition region motion along several field
lines in several `DF-like' events and fitted the motion with
parabolas. The resulting correlation between measured maximum
velocities and decelerations (Fig.~\ref{sim_dec-maxv}) show the same
trend as found in the observed DFs (upper left panel of
Fig.~\ref{fig_scattergood}): large decelerations are correlated with
large maximum velocities. The range in maximal velocities and
decelerations is approximately the same as that found in the observed
DFs, and the slope in the correlation is about the same. Inspection of
Fig.~\ref{sim_xt} also indicates that the simulated DFs have the same
range of heights and durations that the observed DFs display.

Higher resolution 3D simulations with varying magnetic topologies will
be necessary to explain the range of DF widths, the internal DF
substructure, and what determines the sometimes sheet-like structure
of some DFs. Such simulations will also help to fully explain the
range of periods and lifetimes that we observe.

\subsection{Quiet Sun equivalents}
\label{section:qs}

How do these findings relate to quiet Sun phenomena such as mottles
and spicules at the limb? There are many similaties between mottles
and DFs. Both phenomena appear as highly dynamic, dark features in the
wings and core of \halpha. They are associated with magnetic flux
concentrations that can be observed as bright points in the blue wing
of \halpha. More importantly, like DFs, mottles also follow parabolic
paths. In an excellent analysis of the dynamics of inclined mottles
(which they refer to as disk spicules) \citet{Suematsu+etal1995} find
that:
\begin{itemize}
\item mottles seem to be generated several hundred kilometers above
  the photosphere,
\item mottles undergo real mass motion of order 10-30~\kms, extending
  up to 2-10 Mm during their lifetime of 2-15 minutes,
\item most mottles are observed to ascend and descend,
\item mottles in weak plage, or the interior of network cells are
  more vertical and shorter than the more inclined quiet Sun mottles,
\item the appearance of mottles is varied, they can be thick, thin,
  tapered or not, curved or straight,
\item some mottles show downward motion close to their base during the
  ascending phase,
\item the decelerations observed in mottles are too small to be
  consistent with a purely ballistic flight (at solar gravity) and the
  observed Doppler shifts,
\item the largest doppler signal appears at the beginning of the ascending
  phase, and at the end of receding phase,
\item the observed velocity profiles are compatible with impulsive
  acceleration and constant deceleration afterward. 
\item the maximum upward velocity is usually identical in amplitude to
  the maximum downward velocity,
\item there is a linear relationship between the duration and the
  maximum length of mottles,
\item there is a correlation between the average velocity and
  maximum length,
\item nearby mottles are often found to be driven by a common
  disturbance.
\end{itemize}

These findings agree strikingly well with our observations and
modelling of DFs, with only a few discrepancies. The lifetimes and
maximum lengths of the mottles \citet{Suematsu+etal1995} analyzed are
typically longer than those of the DFs we describe. We believe this is
in part a selection effect, as we chose to focus only on highly
dynamic and shorter features in our dataset. We note that our dataset
also contains a large number of highly inclined fibrils that are
longer lived and less dynamic than the DFs (see, e.g.,
Fig.~\ref{fig_osc_panels}). In addition, preliminary analysis of
high-quality SST data of quiet Sun mottles indicates that the most
inclined mottles are on average longer, but that many of the mottles
actually show lengths (1,000-5,000 km) that are similar to those of
DFs. Until further detailed analysis, it can however not be excluded
that DFs associated with plage are on average somewhat shorter in
length and lifetime than quiet Sun mottles. Preliminary analysis of
our numerical simulations suggests that these differences could be
related to large scale differences in magnetic topology. Further work
is necessary to establish the exact relationship between mottles and
DFs.  However, it should be noted that in addition to these striking
similarities with mottles, recent observations of limb spicules
\citep{Christopoulou+etal2001} show clear parabolic paths with
decelerations and maximum velocities of the same order as those
reported for mottles \citep{Suematsu+etal1995} and those we find for
DFs.  Older observations of limb spicules have also shown evidence for
correlations between velocities and maximum lengths
\citep{Rush+Roberts1953}, and between duration and height
\citep{Dizer1952}. These correlations are similar to the ones we find
for DFs. In addition, \citet{Lippincott1957} has reported on
collective behavior of spicules or mottles over several thousand
kilometers.

As discussed also by \citet{Hansteen+etal2006}, these similarities
strongly suggest that chromospheric shock waves cause significant
excursions of the upper chromosphere in both active region and quiet
Sun. Such a scenario was proposed by \citet{DePontieu+etal2004} using
SST and TRACE data.  Further detailed analysis of our numerical
simulations can help shed light on several unresolved issues, such as
the role of reconnection in forming quiet Sun jets, or whether the
observed height of limb spicules can be explained by the mechanism
proposed here.

\citet{Hansteen+etal2006} suggest that it is possible that limb
spicules consist of two populations: jets that are driven by shocks
(as described in this paper), and jets caused by reconnection.  The
latter jets could form a subset that on average is taller than the
shock driven jets, and perhaps be a part of a continuous spectrum of
reconnection jets that includes surges, macrospicules and \halpha\ 
upflow events \citep{Chae+etal1998}. These taller jets may be the
features that have UV counterparts (e.g, observed by SUMER), which
\citet{Wilhelm2000} calls ``spicules''.



\section{Summary}
\label{section:summary}

The combination of high-resolution chromospheric data and advanced
numerical simulations strongly suggests that DFs are formed by
upwardly propagating chromospheric oscillations/waves. These waves are
generated in the convection zone/photosphere as a result of global
p-mode oscillations and convective flows. The magnetic topology acts
as a filter, so that only waves of certain periods can propagate into
and through the chromosphere, forming shocks along the way, up to
regions where H$\alpha$ is formed.

Some of the key observational/modelling results on which this
interpretation is based are:

\begin{itemize}
\item DFs follow a parabolic path
\item DFs in dense plage regions are shorter, slower, undergo
  larger deceleration and live shorter
\item DFs in regions adjacent to dense plage regions are longer,
  faster, undergo less deceleration, and live longer
\item DFs in dense plage regions show periodicities around 180 s
\item DFs in regions adjacent to dense plage regions show
  periodicities around 300 s
\item the velocity and deceleration of all DFs is well correlated:
  the higher the velocity, the more deceleration
\item this correlation is reproduced well in numerical simulations
  where DFs are driven by shocks caused by upward
  propagating waves in the chromosphere.
\item DFs in the simulations follow parabolic paths
  as well.
\item the simulations reproduce the observed values of maximum
  velocities, decelerations and lengths well.
\end{itemize}

From the results describe above and by \citet{Hansteen+etal2006}, it
seems clear that in active regions, most dynamic fibrils are formed by
chromospheric shocks driven by convective flows and oscillations in
the photosphere.



\acknowledgements{BDP was supported by NASA grants NAG5-11917,
  NNG04-GC08G and NAS5-38099 (TRACE), and thanks the ITA/Oslo group
  for excellent hospitality. VHH thanks LMSAL for excellent
  hospitality during the spring of 2006. This research was supported
  by the European Community's Human Potential Programme through the
  European Solar Magnetism Network (ESMN, contract HPRN-CT-2002-00313)
  and the Theory, Observation and Simulation of Turbulence in Space
  (TOSTISP, contract HPRN-CT-2002-00310) programs, by The Research
  Council of Norway through grant 146467/420 and through grants of
  computing time from the Programme for Supercomputing. The Swedish
  1-m Solar Telescope is operated on the island of La Palma by the
  Institute for Solar Physics of the Royal Swedish Academy of Sciences
  in the Spanish Observatorio del Roque de los Muchachos of the
  Instituto de Astrof{\'\i}sica de Canarias. The authors thank K.
  Schrijver and M.  DeRosa for useful discussions and comments. }

\clearpage

\clearpage

\end{document}